%% file: Main.tex
\documentclass[%
superscriptaddress,
preprint,
 amsmath,amssymb,
 aps,
]{revtex4-2}

\usepackage{graphicx}
\usepackage{dcolumn}
\usepackage{bm}

\usepackage{appendix}
\usepackage{array}
\usepackage{tikz}
\usepackage{booktabs}
\usepackage{caption}
\usepackage{subcaption}
\usepackage{cancel}

\input{newcommands}

\begin{document}

\preprint{APS/123-QED}

\title{
Energy-Preserving Coupling of Explicit Particle-In-Cell with Monte Carlo Collisions
}

\author{Jean-Luc Vay}
 \email{jlvay@lbl.gov}
 \affiliation{Lawrence Berkeley National Laboratory, Berkeley, CA, 94720, USA}
\author{Justin Ray Angus}%
 \affiliation{Lawrence Livermore National Laboratory, Livermore, CA, 94551, USA}
\author{Olga Shapoval}
 \affiliation{Lawrence Berkeley National Laboratory, Berkeley, CA, 94720, USA}
\author{R\'emi Lehe}
 \affiliation{Lawrence Berkeley National Laboratory, Berkeley, CA, 94720, USA}
\author{David Grote}%
 \affiliation{Lawrence Livermore National Laboratory, Livermore, CA, 94551, USA}
\author{Axel Huebl}
 \affiliation{Lawrence Berkeley National Laboratory, Berkeley, CA, 94720, USA}



\date{\today}

\begin{abstract}
The Particle-In-Cell (PIC) and Monte Carlo Collisions (MCC) methods are workhorses of many numerical simulations of physical systems. Recently, it was pointed out that, while the two methods can be exactly - or nearly - energy-conserving independently, combining the two is leading to anomalous numerical heating. 
This paper reviews the standard explicit PIC-MCC algorithm, elucidates the origins of the anomalous numerical heating and explains how to couple the two methods such that the anomalous numerical heating is avoided. 

\end{abstract}

\maketitle


\section{\label{sec:intro}Introduction}
The Particle-In-Cell and Monte Carlo methods are workhorses of many numerical simulations of physical systems. The two are combined within the Particle-In-Cell Monte Carlo Collision (or PIC-MCC) method, formalized by Birdsall in a 1991 review paper~\cite{BirdsallIEEE1991}, and adopted or studied by many (\cite{BirdsallIEEE1991} cited by over 1,000 papers). Recently, it was pointed out that, while the two methods can be energy-conserving independently, combining the two is leading to anomalous numerical heating~\cite{AlvesPRE2021,AngusJCP2022}.
Mitigations were proposed, including using low-pass filtering~\cite{AlvesPRE2021}, while it was shown that energy is exactly conserved when coupling MCC with an energy-conserving implicit PIC method~\cite{AngusJCP2022}.

In this paper, we review the standard explicit PIC-MCC algorithm, show that the cause of numerical heating is due to the breaking of time-centering of the standard leapfrog loop that is at the heart of the PIC method, and that restoring time-centering ensures better energy conservation and therefore removes the anomalous heating reported earlier~\cite{AlvesPRE2021,AngusJCP2022}. 
We also show that this method applies to both electrostatic and electromagnetic PIC, suppressing anomalous heating in both cases.

The paper is organized as follows. The standard PIC-MCC algorithm is reviewed in Section~\ref{sec:standard}, together with the consequences on energy conservation.
Section~\ref{sec:novel} introduces a time-centered PIC-MCC algorithm and explores its effect in test simulations. In section~\ref{sec:Analysis}, we use numerical analysis to explain the properties of the proposed algorithm. We first elucidates its fundamental properties with a simplified subset that explores the motion of a single particle experiencing harmonic oscillations with random rotations as a proxy for collisions, and then generalize this analysis to  the full PIC-MCC cycle. Section~\ref{sec:implementation} discusses the practical implementation of this algorithm in a PIC code.

\section{\label{sec:standard}Standard PIC-MCC and consequences}
\begin{figure}
    \centering
     \begin{subfigure}[b]{0.8\textwidth}
\caption{Standard PIC loop with MCC before or after the velocity push}
\includegraphics[width=0.7\linewidth]{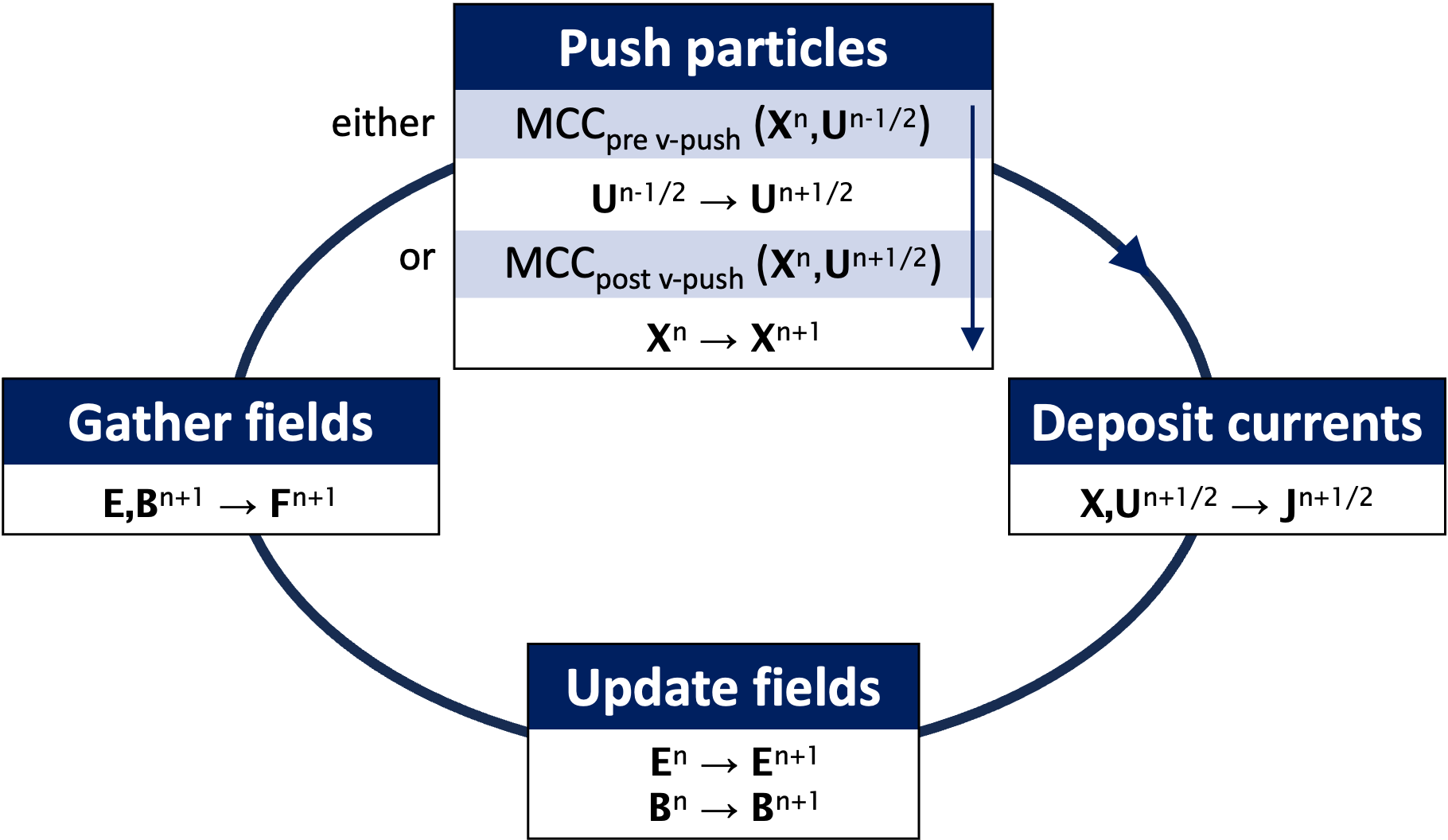}
\label{fig:PIC-MCC} 
\end{subfigure}
     \centering
     \begin{subfigure}[b]{0.8\textwidth}
         \centering
         \caption{Novel PIC loop with MCC in the middle of the velocity or the position push}
\includegraphics[width=0.7\linewidth]{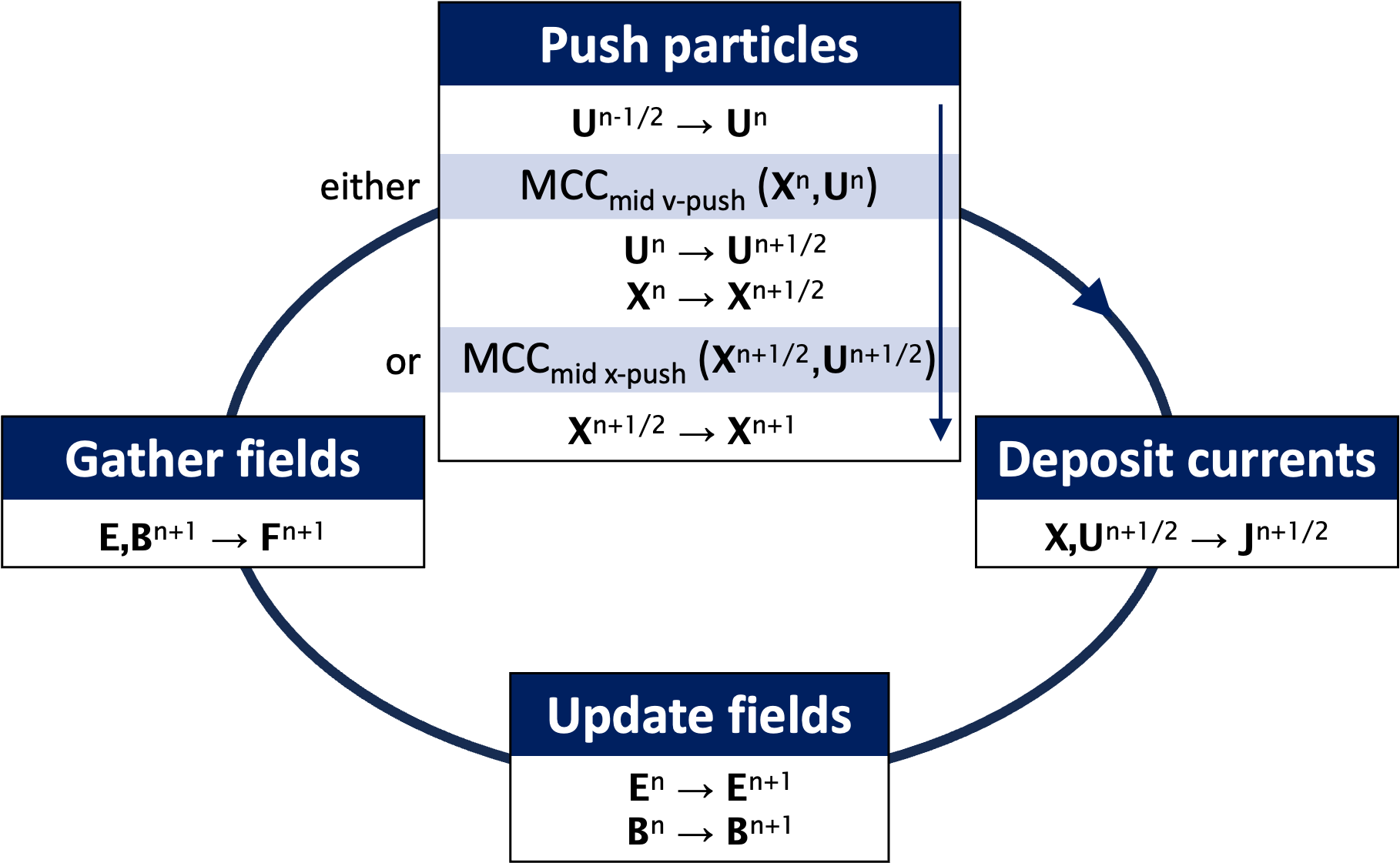}
\end{subfigure}
\caption{Particle-In-Cell loop with Monte Carlo Collisions (MCC). The operations occur sequentially clockwise around the loop, and from top to bottom in the ``Push particles'' inset. In the standard configuration (top), the MCC step usually occurs either before or after pushing the particles velocities. In the proposed configuration (bottom), the MCC step occurs either in the middle of the velocities push or in the middle of the positions push.}
\label{fig:PIC-MCCnew} 
\end{figure}

In the standard PIC-MCC loop, depicted in Fig.~\ref{fig:PIC-MCC}, the collisions occur either before (``pre v-push'') or after (``post v-push'') the particles velocity update. In any of these two cases, collisions occur when positions and velocities are shifted in time, due to the leapfrogging of positions and velocities updates, which, as will be shown, is the source of the anomalous heating reported earlier~\cite{AlvesPRE2021,AngusJCP2022} and explained here.

Similarly to previous work~\cite{AlvesPRE2021,AngusJCP2022}, this paper analyses the evolution of a 2-D slab with a relativistic PIC code (using either an electromagnetic or electrostatic field solver) of a uniform electron-proton plasma initialized with the parameters listed below and summarized in Table ~\ref{tab:2d_test_parameters}).
The plasma of density $n_0=n_e=n_i=10^{30}\ m^{-3}$ is initialized at thermal equilibrium $T_0=T_e=T_i=100 \ eV$,  filling the periodic simulation box of size $10 \delta_e \times  10 \delta_e$. Here,  $\delta_e = c\omega_{pe}^{-1}$ is the skin depth, and $\omega_{pe} = \sqrt{n_0e^2 / m_e \varepsilon_0}$ is the electron plasma frequency. The temporal and spatial resolutions are set to $\omega_{pe} \Delta t = 0.1$ and $\Delta x = \Delta z = 0.25 \delta_e$, respectively. The simulation is initialized with $N_{ppc}=100$ macroparticles per cells (for each species). 
\begin{table}
\centering
\begin{tabular}{l|c}
\midrule
\midrule
Plasma density & $n_0 = 10^{30}$ $m^{-3}$ \\
Simulation box & $L_x  = L_z = 10\delta_e$ \\
Number of grid points & $N_x = N_z = 40$ \\
Spatial resolution & $\Delta x = \Delta z = 0.25 \delta_e$ \\
Timestep & $\Delta t = 0.1 \omega_{pe}^{-1}$ \\
Number or macroparticles per cell & $100$ \\
Order of the shape factor & 2 \\
\hline
\end{tabular}
\caption{\label{tab:2d_test_parameters}Parameters of the 2D energy conservation test simulations.}
\end{table}

\begin{figure}
     \centering
     \begin{subfigure}[b]{1.0\textwidth}
         \centering
         \caption{Standard electrostatic PIC without (left) and with (right) MCC}
         \includegraphics[width=0.49\textwidth]{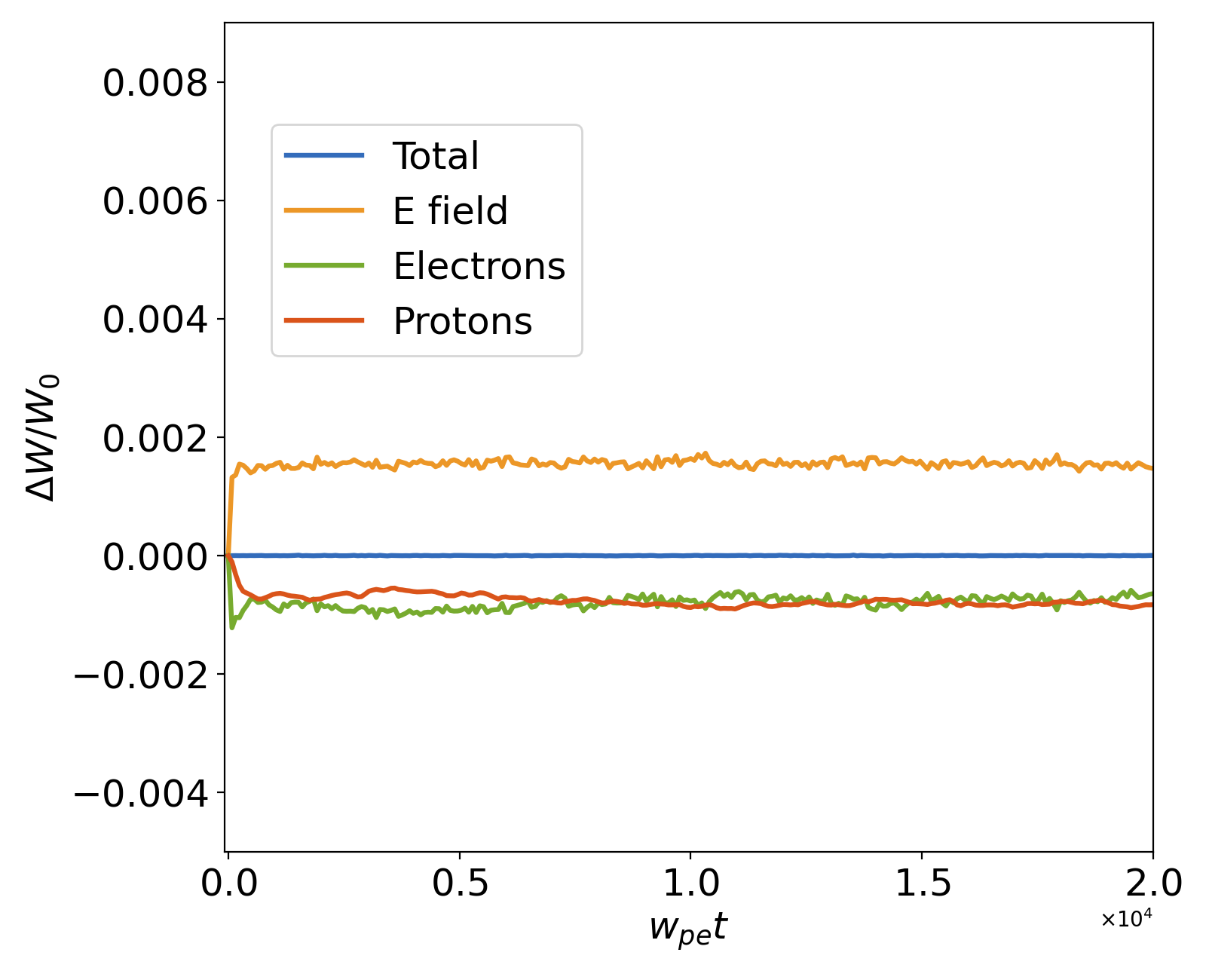} 
         \includegraphics[width=0.49\textwidth]{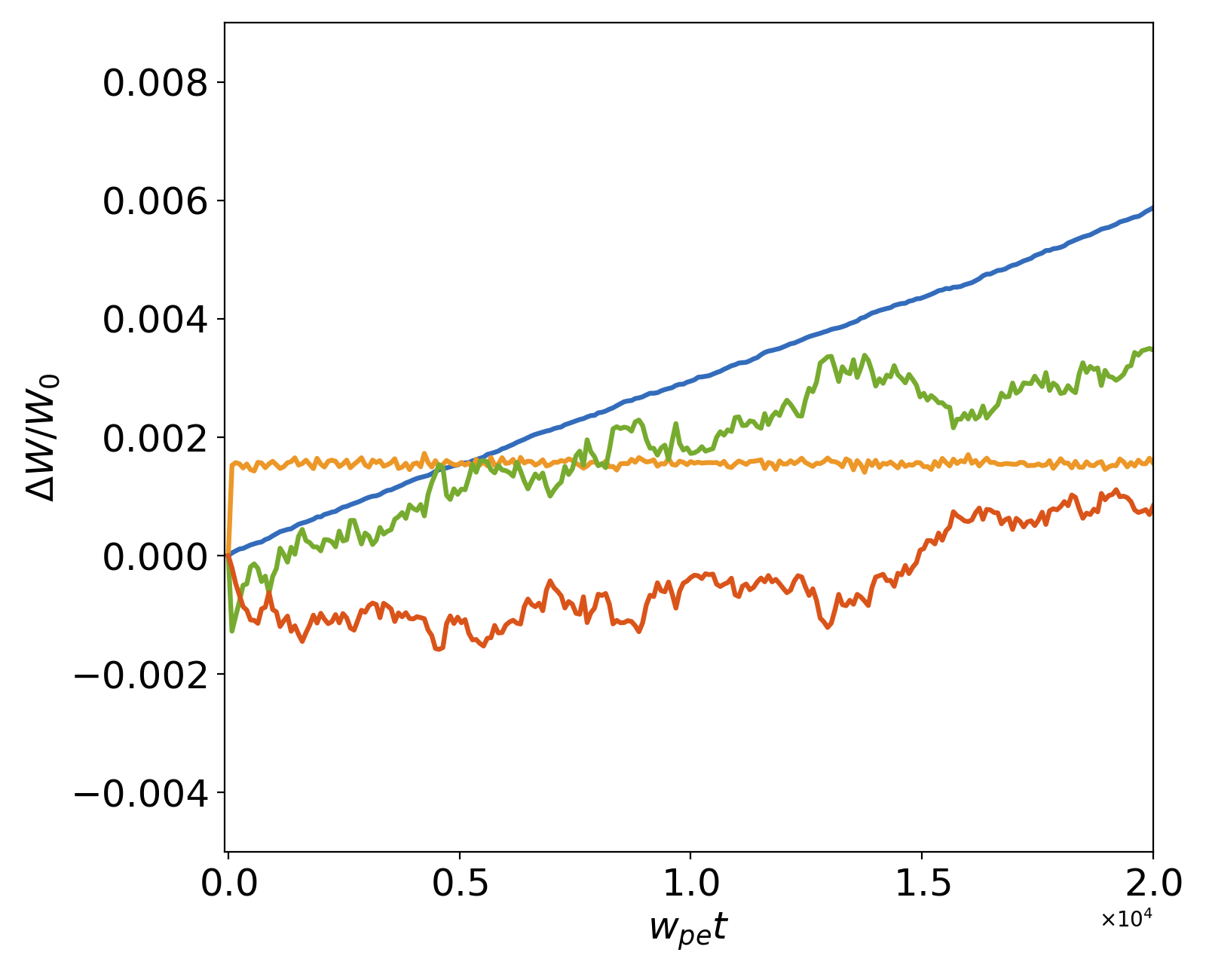} 
         \label{fig:es_energy_before_v_push}
     \end{subfigure}
     \hfill
     \begin{subfigure}[b]{1.0\textwidth}
         \centering
         \caption{Standard electromagnetic PIC without (left) and with (right) MCC}
         \includegraphics[width=0.49\textwidth]{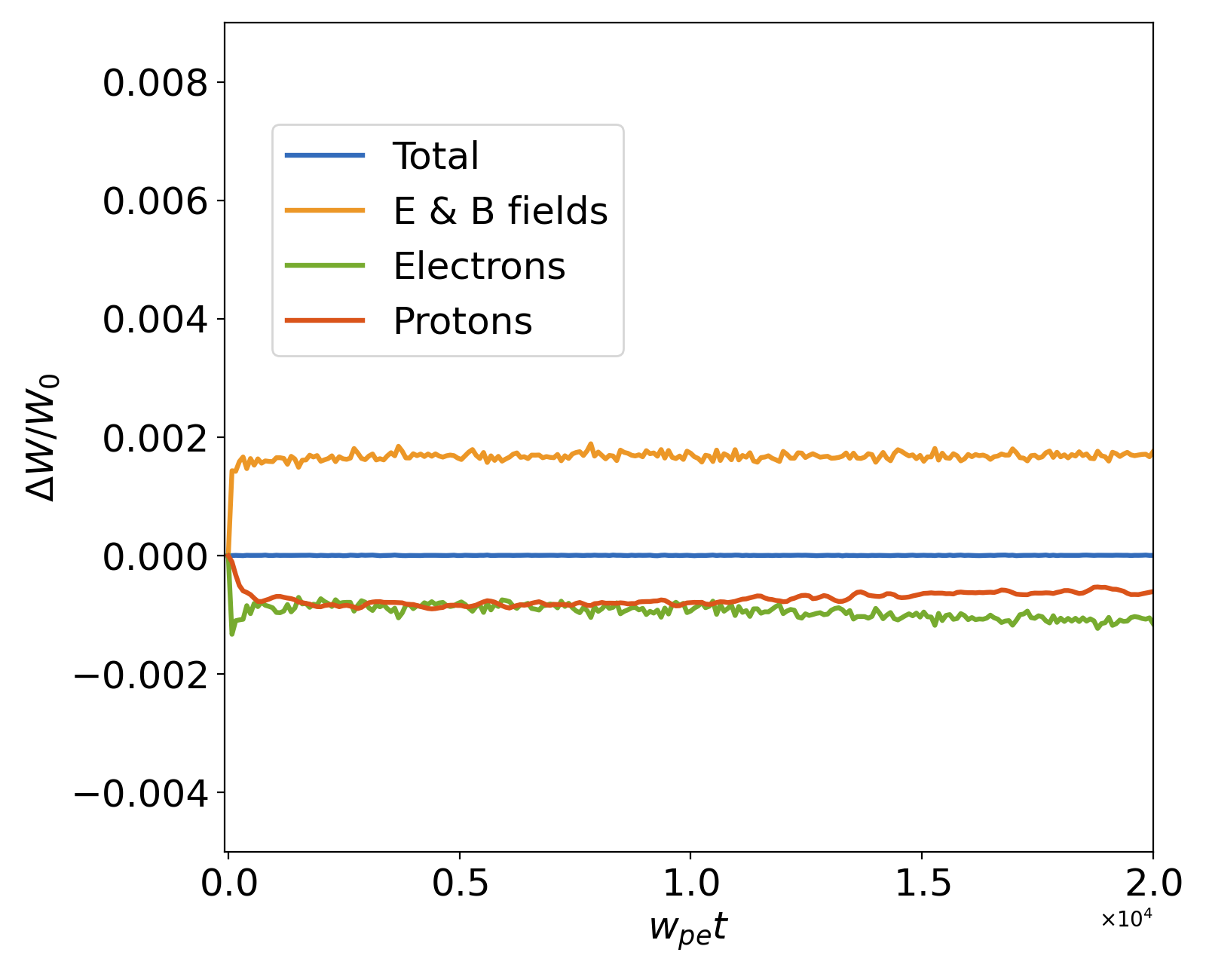} 
         \includegraphics[width=0.49\textwidth]{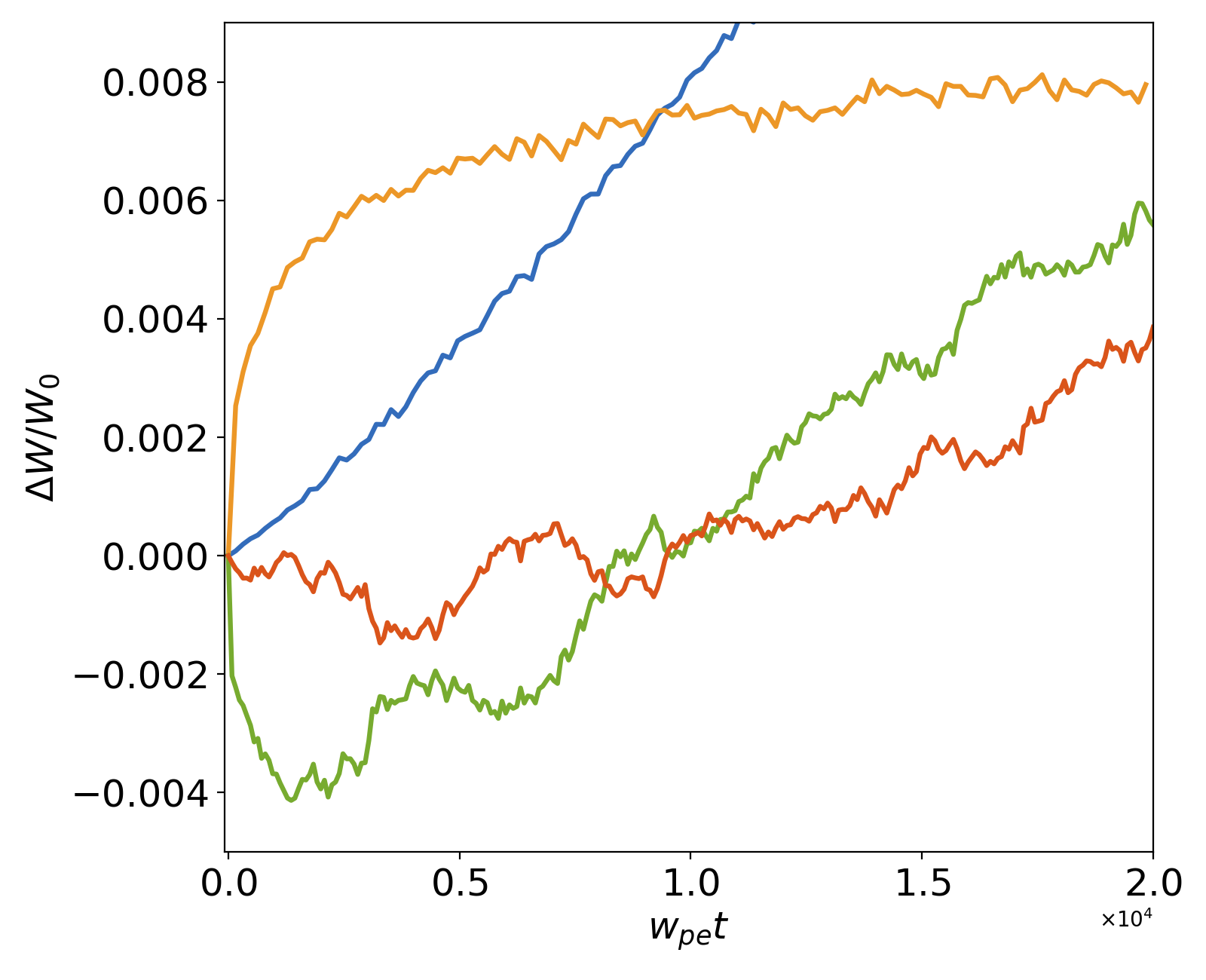} 
         \label{fig:em_energy_before_v_push}
     \end{subfigure}
     \hfill
        \caption{The relative change  $\Delta W / W_0$ of energy as a function of time $\omega_{pe} t$ from 2D simulations of the numerical energy conservation test case using standard (a) electrostatic and (b) electromagnetic PIC (left) without MCC and (right) with MCC before the velocity push (pre v-push). Without collisions (left), the total energy (blue) is conserved to a high level of precision. With collisions (right), the total energy exhibits a secular growth.
        }
        \label{fig:energy_standard_pic_mcc}
\end{figure}

As reported in earlier work~\cite{AlvesPRE2021,AngusJCP2022}, even when the PIC loop uses the so-called `energy-conserving' (also known as Galerkin) algorithm to gather the electric (and magnetic) fields onto macroparticles~\cite{Birdsalllangdon} (which conserves energy exactly at the limit of infinitesimal time steps) and even though the MCC module conserves energy exactly, increased numerical heating occurs when combining PIC and MCC together in such a loop. While this was observed and analyzed only in the context of electromagnetic (EM) PIC-MCC~\cite{AlvesPRE2021,AngusJCP2022}, it is also occurring with electrostatic (ES) PIC-MCC, as shown in Fig.~\ref{fig:energy_standard_pic_mcc}, where the evolution of energy from the fields, electrons, protons and total is plotted (normalized to the initial total energy) using either an electrostatic solver (top) or an electromagnetic (Yee) solver (bottom), without (left) and with (right) Monte Carlo Collisions. In either case, electrostatic and electromagnetic PIC, the standard coupling of PIC and MCC (placing MCC before the velocity push for the tests reported here) leads to anomalous numerical heating. The observation with the electromagnetic PIC solver are in agreement with previous results~\cite{AlvesPRE2021,AngusJCP2022} while the result using the electrostatic PIC solver suggests that the explanations given previously~\cite{AlvesPRE2021} that the numerical heating is due to an interplay of electromagnetic radiations and particle motion is either incomplete or inaccurate. In the remainder of the paper, we propose and analyze a new algorithm that does not lead to the anomalous heating and also elucidate the underlying causes. 

\section{\label{sec:novel}Novel PIC-MCC algorithm}

\subsection{\label{subsec:algorithm}The algorithm}

With the proposed algorithm, the collision now occurs in the middle of the velocity push (``mid v-push'') or the position push (``mid x-push''), as depicted in Fig.~\ref{fig:PIC-MCCnew}. More specifically, the PIC-MCC loop that is proposed includes either a velocity push that is split in the following series of steps: 
\begin{align}
    \ub^n_p &= \ub^{n-1/2}_p + \frac{q_p}{m_p} (\Eb^n + \vb^n_p \times \Bb^n) \Delta t/2 && \text{push velocities by half a time step} \\
    \ub^{n}_p &\rightarrow \tilde{\ub}^{n}_p && \text{perform collisions} \\
    \ub^{n+1/2}_p &= \tilde{\ub}^{n}_p + \frac{q_p}{m_p} (\Eb^n + \tilde{\vb}^n_p \times \Bb^n) \Delta t/2 && \text{push velocities by half a time step}
\end{align}
or a position push that is split in the following series of steps:
\begin{align}
    \xb^{n+1/2}_p &= \xb^{n}_p +  \ub^{n+1/2}_p/\gamma_p \Delta t/2 &&\text{push positions by half a time step} \\
    \ub^{n+1/2}_p &\rightarrow \tilde{\ub}^{n+1/2}_p &&\text{perform collisions} \\
    \xb^{n+1}_p &= \xb^{n+1/2}_p +  \tilde{\ub}^{n+1/2}_p/\tilde{\gamma_p} \Delta t/2 &&\text{push positions by half a time step} 
\end{align}

In each case, the rest of the PIC loop is left unchanged, except for the current deposition in the latter (mid x-push) case that must take into account the velocity change occurring in the middle of the position push.
It is important to
note that the definition of the velocity used for the $\vb\times \Bb$ term depends on the algorithm of the particle pusher~\cite{Vaypop2008,Higuera2017} and is generally not simply given by $\ub=\gamma\vb$.
This is discussed further when discussing practical implementation in a PIC code in Sec~\ref{sec:implementation}.

\subsection{\label{subsec:fullPIC_tests}Numerical examples}

Two examples are considered in this section, starting with the 2-D uniform plasma test reported in Section~\ref{sec:standard} and a 1-D magnetic-driven piston collisional shock.

\subsubsection{\label{subsubsec:fullPIC_tests2D}2-D uniform plasma simulations}
Simulations were performed using the same 2D uniform plasma setup as in section~\ref{sec:standard}, using the mid v-push and the mid x-push PIC-MCC algorithms with both the electrostatic and the electromagnetic PIC methods. 
In all cases, the use of the proposed algorithms lead to excellent energy conservation without exhibiting the spurious energy growth that is observed otherwise in the literature~\cite{AlvesPRE2021,AngusJCP2022} and in Figure~\ref{fig:energy_standard_pic_mcc}. 
This observation is further explained with numerical analysis in Section~\ref{sec:Analysis}.

\begin{figure}
     \centering
     \begin{subfigure}[b]{1.\textwidth}
         \centering
         \caption{Novel mid v-push (left) electrostatic and (right) electromagnetic PIC-MCC}
         \includegraphics[width=0.45\textwidth]{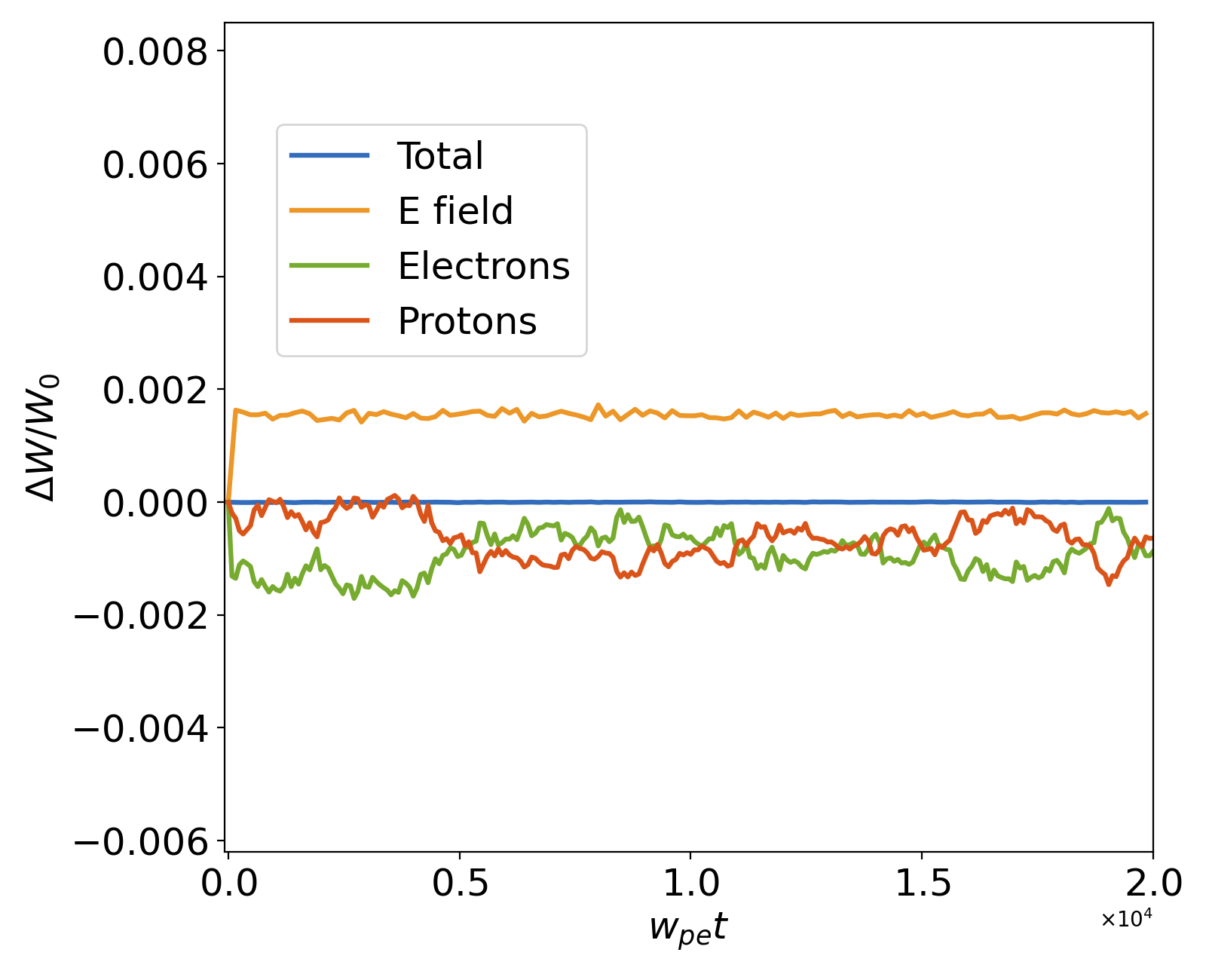}
         \includegraphics[width=0.45\textwidth]{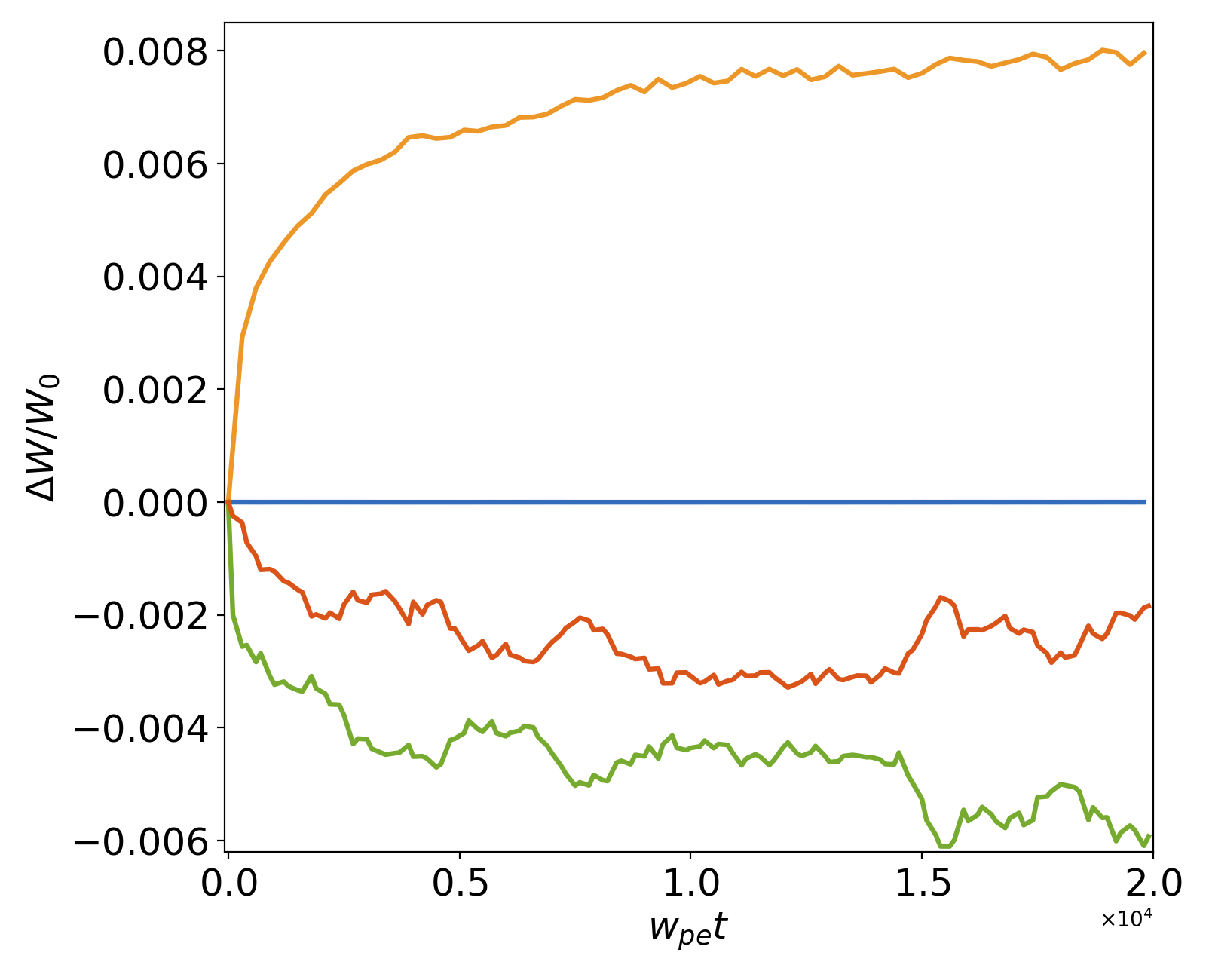}
         \label{fig:three sin x}
     \end{subfigure}
     \hfill
     \begin{subfigure}[b]{1.\textwidth}
         \centering
         \caption{Novel mid x-push (left) electrostatic and (right) electromagnetic PIC-MCC}
         \includegraphics[width=0.45\textwidth]{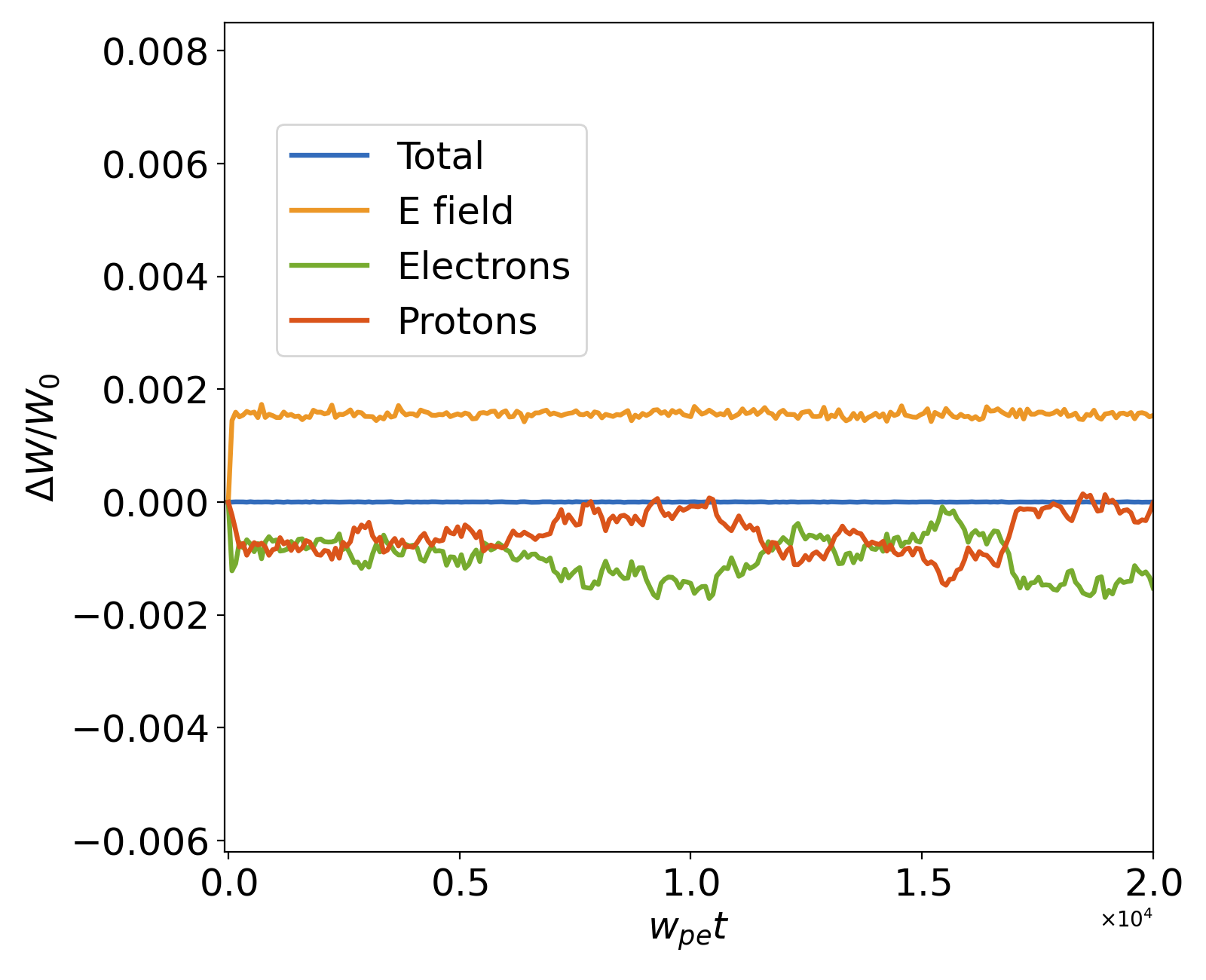}
         \includegraphics[width=0.45\textwidth]{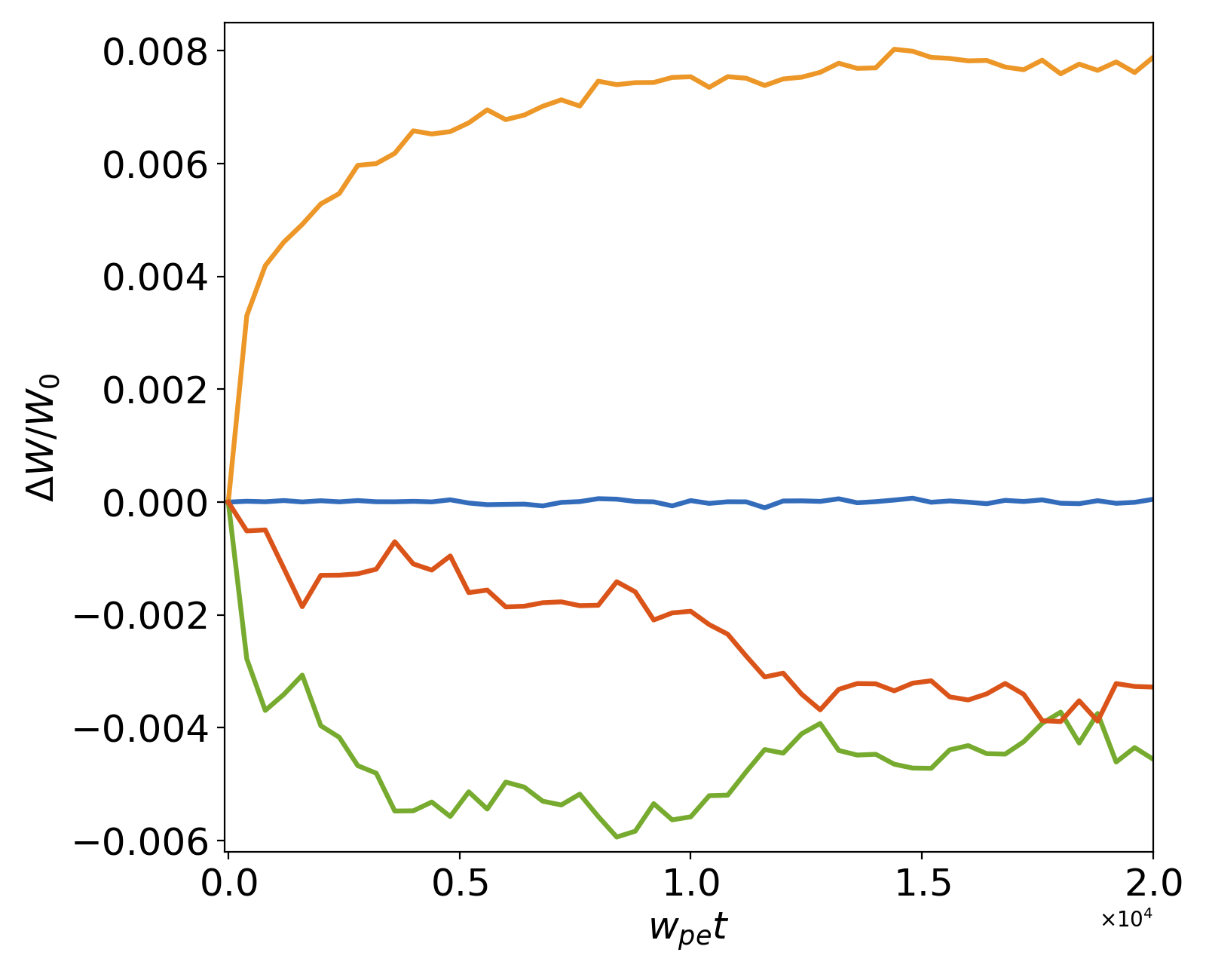}
     \end{subfigure}
     \hfill
        \caption{The relative change  $\Delta W / W_0$ of energy as a function of time $\omega_{pe} t$ from 2D simulations of the numerical energy conservation test case using novel (a) ES and (b) EM PIC with MCC model in the middle of (top) the velocity push and (bottom) the position push. While the total energy (blue) is not conserved to machine precision with full explicit PIC, it is well conserved with the new placement of MCC in the middle of the velocity or position push (compare with the secular numerical heating observed in Figure~\ref{fig:energy_standard_pic_mcc}).}
        \label{fig:three graphs}
\end{figure}

\subsubsection{\label{subsubsec:fullPIC_tests_Piston}1-D magnetic-driven piston collisional shock simulations}
\begin{figure}
    \centering
    \includegraphics[scale=0.55]{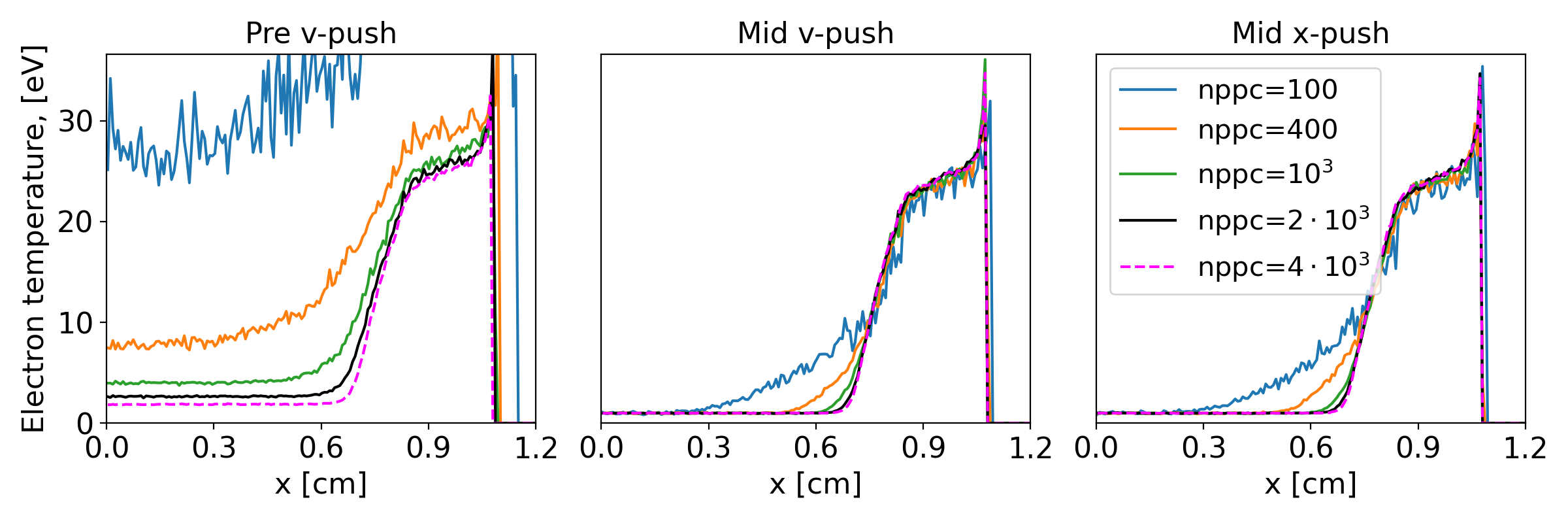}
    \caption{The electron temperature at $t=66$ ns from 1D magnetic-piston simulations, obtained using the WarpX electromagnetic PIC solver with different MCC placements: (left) pre v-push MCC, (middle) mid v-push MCC and (right) post v-push MCC. Several simulations were run, raising the number of macroparticles per cell from $nppc=100$ (blue), 400 (orange), 1000 (green), 2000 (black) and 4000 (magenta).}
    \label{fig:magnetic_piston}
\end{figure}

In this section, the mid v-push and mid x-push PIC-MCC algorithms are tested and compared to the standard pre v-push one on the 1-D magnetic-piston driven collisional shock case that was reported in Ref.~\cite{AngusJCP2023}.
In these simulations, a fully ionized deuterium plasma slab of length $L_x=1.54$ cm is initialized with uniform initial temperature $T_0=T_e=T_i=1 eV$ and density $n_0 = n_e=n_i=10^{23}\ m^{-3}$. The temporal and spatial resolutions are set to $ \Delta t = 20~fs \ (\omega_{pe} \Delta t = 0.36)$ and $N_x=216$. A lower boundary is treated as a symmetry plane, while at the upper boundary an external magnetic field is applied, rising linearly from $0$ to a peak value of $B_0 = 2.667$ T over $8.154$ ns and remaining constant for later times. 
The magnetic field acts as a piston, creating a shock propagating from the upper toward the lower boundary.
The simulations are initialized with different number of macroparticles $N_{ppc}$ per cells (for each species), varying between 100 and 4000. 
Snapshots of the electron temperature at $t=66$~ns are shown in Fig.~\ref{fig:magnetic_piston}.

Consistent with the results reported in \cite{AngusJCP2023}, using the standard PIC-MCC algorithm with pre v-push MCC leads to significant noise and anomalous temperature increase in the entire slab. 
While the anomalous temperature increase is reduced by increasing the number of macroparticle per cell, the convergence rate is very slow, rendering converged simulations prohibitively expensive computationally. 

By contrast, using the proposed PIC-MCC algorithm with either the mid v-push or the mid x-push placement of MCC leads to much reduced anomalous numerical heating with a much faster convergence rate when increasing the number of macroparticles. 
It was observed in~\cite{AngusJCP2023} that a remaining numerical heating is observed with implicit PIC-MCC and is reduced when using larger time steps. 
Tests that will be reported in a future paper show that the level of remaining numerical heating is at the same level whether using implicit PIC-MCC or explicit PIC-MCC with mid v-push or mid x-push centering, suggesting the same origin. 
Further studies are needed to fully elucidate the origins of this remaining numerical heating, which go beyond the scope of this paper. 

\section{\label{sec:Analysis}Analysis of the properties the proposed PIC-MCC algorithm}

Analyses of the properties of the proposed PIC-MCC algorithm are reported in this section, starting with the analysis of a single-particle harmonic oscillator model, followed by an analysis of the full PIC-MCC loop with an ensemble of macroparticles.

\subsection{\label{subsec:HO}Elucidation of fundamental properties with a single-particle harmonic oscillator model}

In this subsection, a simplified subset of the full PIC-MCC algorithm is considered, so as to elucidate the fundamental properties and the logic behind the proposed algorithm. In particular, we show that the same effect (i.e., conservation of energy only when the MCC module is placed in the middle of the x-push or v-push) also occurs for a single-particle model, and is thus not due to collective interactions.

Noting that the modifications of the algorithm occur in the particle pusher portion of the PIC loop, the simplified analysis will focus on the particle pusher only using the harmonic oscillator model, widely used in the community for such analyses~\cite{Birdsalllangdon,HockneyEastwood1988}. 
To further simplify, a single particle is considered and collisions are replaced by a rotation of the velocity vector of the particle by a random angle, the simplest of operation that still retains the property of energy conservation of MCC. We refer to this operation as a \emph{transverse velocity kick} in the rest of this section.
A complete analysis of the full PIC loop, which includes charge and current deposition, field solve and gathering of the electromagnetic fields from the grids onto the macroparticles and Coulomb collisions on an ensemble of macroparticles, is presented in the next section.

Assuming the non-relativistic limit for simplicity, the leapfrog integration of a single particle experiencing harmonic oscillations is
\begin{align}
    \vb^{n+1/2} - \vb^{n-1/2} &= \frac{q}{m} \Eb^n \Delta t, \label{Eq:HOV}\\
    \xb^{n+1} - \xb^n &= \vb^{n+1/2} \Delta t, \label{Eq:HOX}\\
    \Eb^{n+1} &= - \kappa \xb^{n+1}, \label{Eq:HOE}
\end{align}
where $\kappa$ is a constant. 

To simplify the description of the placement of collisions (here transverse velocity kicks) in the leapfrog loop, the system (\ref{Eq:HOV}-\ref{Eq:HOE}) is rewritten using the velocity Verlet split, as an integrator that updates each components $x$ and $v$ from time step $n$ to $n+1$:
\begin{align}
    \vb^{n+1/2} &= \vb^{n} - \frac{\kappa q}{2m} \xb^n \Delta t, & \text{(step 1)} \label{Eq:Verlet1}\\
    \xb^{n+1/2} &= \xb^n + \vb^{n+1/2} \Delta t/2, & \text{(step 2)} \label{Eq:Verlet2}\\
    \xb^{n+1} &= \xb^{n+1/2} + \vb^{n+1/2} \Delta t/2, & \text{(step 3)} \label{Eq:Verlet3}\\
    \vb^{n+1} &= \vb^{n+1/2} - \frac{\kappa q}{2m} \xb^{n+1} \Delta t. & \text{(step 4) \label{Eq:Verlet4}}
\end{align}
Using the standard PIC-MCC method, the transverse velocity kick is typically inserted between step 1 and 2 or between step 3 and 4 (i.e., respectively after or before the velocity push), while with the new method, it is inserted either in the middle of the velocity push (i.e., before step 1 or, equivalently, after step 4) or in the middle of the position push (i.e., between steps 2 and 3).

Without ``collisions'', four simple expressions of the discrete energy can be identified that lead to exact energy conservation over time:
\begin{align}
    W^{n+1/2}_{(1)} &= \frac{1}{2}m (\vb^{n+1/2})^2 + \frac{\kappa q}{2} \xb^n \xb^{n+1}, \label{Eq:WV1/2}\\
    W^{n}_{(2)} &= \frac{1}{2}m \vb^{n-1/2} \vb^{n+1/2} + \frac{\kappa q}{2} (\xb^{n})^2, \label{Eq:WX}\\ 
    W^{n+1/2}_{(3)} &= \frac{1}{2}m \vb^n \vb^{n+1} + \frac{\kappa q}{2} (\xb^{n+1/2})^2, \label{Eq:WX1/2} \\
    W^{n}_{(4)} &= \frac{1}{2}m (\vb^{n})^2 + \frac{\kappa q}{2} \xb^{n-1/2} \xb^{n+1/2}. \label{Eq:WV}
\end{align}
where $\xb^{n+1/2} \equiv (\xb^n+\xb^{n+1})/2$ and $\vb^{n}\equiv(\vb^{n-1/2}+\vb^{n+1/2})/2$, as demonstrated in  Appendix~\ref{subsec:Appendix_HO_energy_conservation_nocollision}.
Any of these four expressions of discrete energy can be used to monitor the energy and verify conservation to machine precision without collisions.

With ``collisions'', one finds that, provided that the transverse velocity kicks conserve energy, only $W^{n+1/2}_{(1)}$ (Eq.~(\ref{Eq:WV1/2})) conserves the energy exactly when transverse velocity kicks are inserted before step (1), while only $W^{n}_{(4)}$ (Eq.~(\ref{Eq:WV})) conserves the energy exactly when transverse velocity kicks are inserted between steps (2) and (3), as demonstrated in Appendices~\ref{subsec:Appendix_HO_energy_conservation_midvpush} and~\ref{subsec:Appendix_HO_energy_conservation_midxpush}.
The fact that at least one of the four expressions is exactly conserved when transverse velocity kicks are placed before step (1) or between steps (2) and (3) proves energy conservation in these cases.
Note that the energy given by the expressions that do not conserve energy with kicks will simply exhibit bumps or dips when a random transverse velocity kick occurs and will still conserve energy exactly between the kicks.

On the contrary, it was found analytically (not reported for brevity) that none of the four expressions given by Eqs.~(\ref{Eq:Verlet1}-\ref{Eq:Verlet4}) conserve energy exactly when the transverse velocity kicks are inserted between step (1) and (2), or between step (3) and (4). 
While this does not prove that the energy will necessarily diverge with time, as observed with the full PIC simulations in Section~\ref{sec:standard} and elsewhere~\cite{AlvesPRE2021,AngusJCP2022}, it is consistent with the observations and with the additional tests reported below.

\subsubsection{\label{subsubsec:HO_tests}Numerical tests}
\begin{figure}
\includegraphics[width=0.95\linewidth]{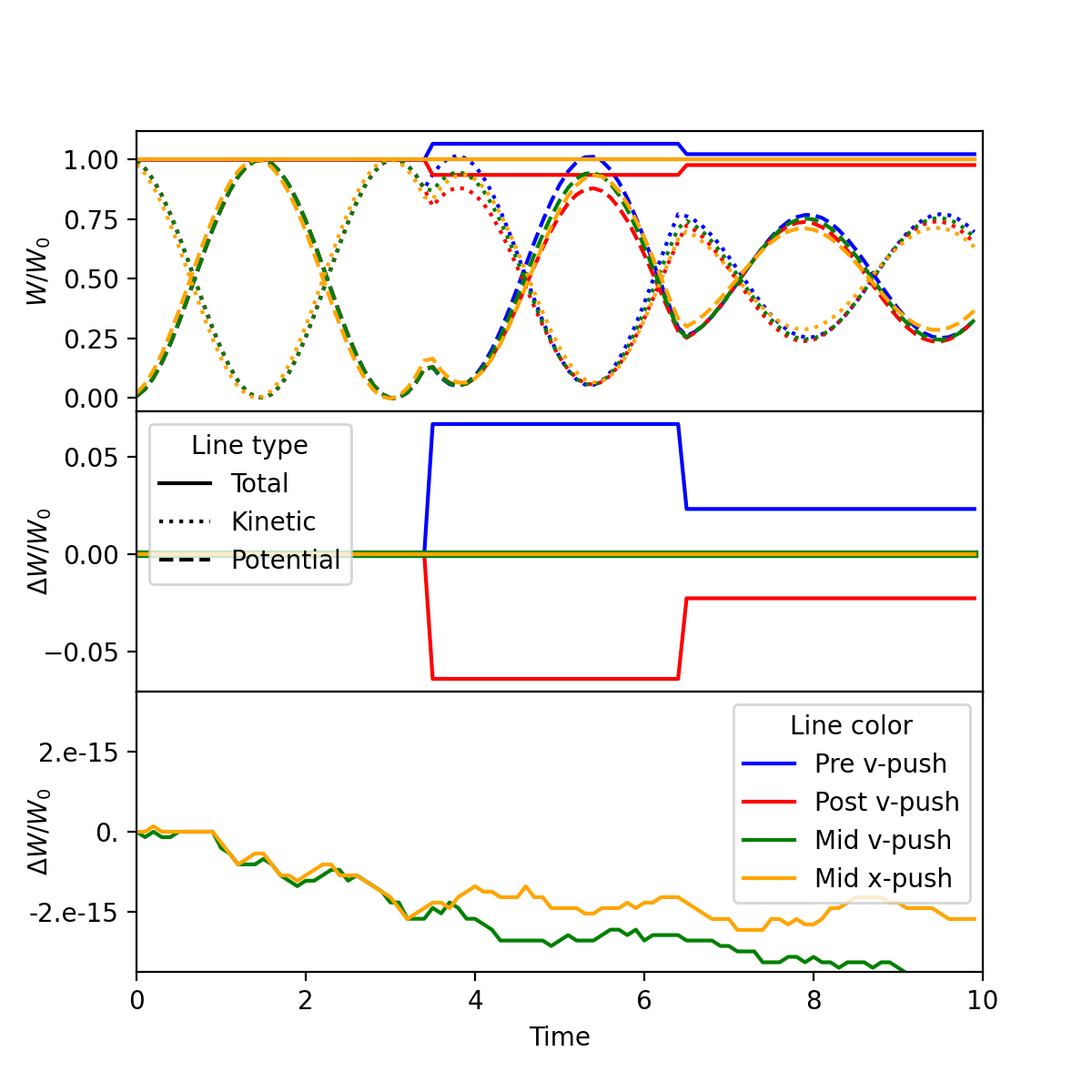}
\caption{(top) Relative total (solid), kinetic (dotted) and potential (dashed) energies for one particle in a 3D harmonic oscillator experiencing two transverse velocity kicks at steps 35 and 65, using the PIC-MCC algorithm with either the MCC placed before the velocity push (blue), after (red) or the new option of placing the MCC in the middle of the velocity push (green) or position push (orange). (middle) Relative total energy deviation for the four options. (bottom) Relative total energy deviation for the new mid v-push (green) and mid x-push (orange) options.
Eq.~(\ref{Eq:WV1/2}):$W^{n+1/2}_{(1)}$ was used to compute the energy for the pre, post and mid v-push algorithms. 
Eq.~(\ref{Eq:WV}):$W^{n}_{(4)}$ was used to compute the energy for the mid x-push algorithm. }
\label{fig:HO} 
\end{figure}

The theory that was exposed above was tested with one particle in a 3D harmonic oscillator following Eqs.(\ref{Eq:Verlet1}-\ref{Eq:Verlet4}) with MCC (transverse velocity kicks) placed before (pre v-push), after (post v-push) or in the middle of the velocity push (mid v-push), or in the middle of the position push (mid x-push). 
Normalized quantities were chosen for simplicity: $\kappa=q=m=1$.
Results are shown in Fig.~\ref{fig:HO} for a particle initialized at $\xb=\{0.,0.,0.\}$ and a finite velocity $\vb=\{0.1,0.,0.\}$.  
Two transverse velocity kicks occur, at time steps 35 and 65, respectively. 
Eq.~(\ref{Eq:WV1/2}):$W^{n+1/2}_{(1)}$ was used to compute the energy for the pre, post and mid v-push algorithms. 
Eq.~(\ref{Eq:WV}):$W^{n}_{(4)}$ was used to compute the energy for the mid x-push algorithm. 
With placement of the MCC before or after the velocity push, the total energy either increases or decreases, while it remains constant to machine precision with the new placement in the middle of the velocity or the position push.
It is important to note that before or after each transverse velocity kick, the energy is conserved to machine precision using all four of Eqs.~(\ref{Eq:Verlet1}-\ref{Eq:Verlet4}), meaning that the deviations from the initial energy observed with the pre v-push and post v-push placements of the random transverse velocity kicks do correspond to actual jumps of the total energy of the system and not to diagnostic approximations. 

\begin{figure}
\includegraphics[width=0.9\linewidth]{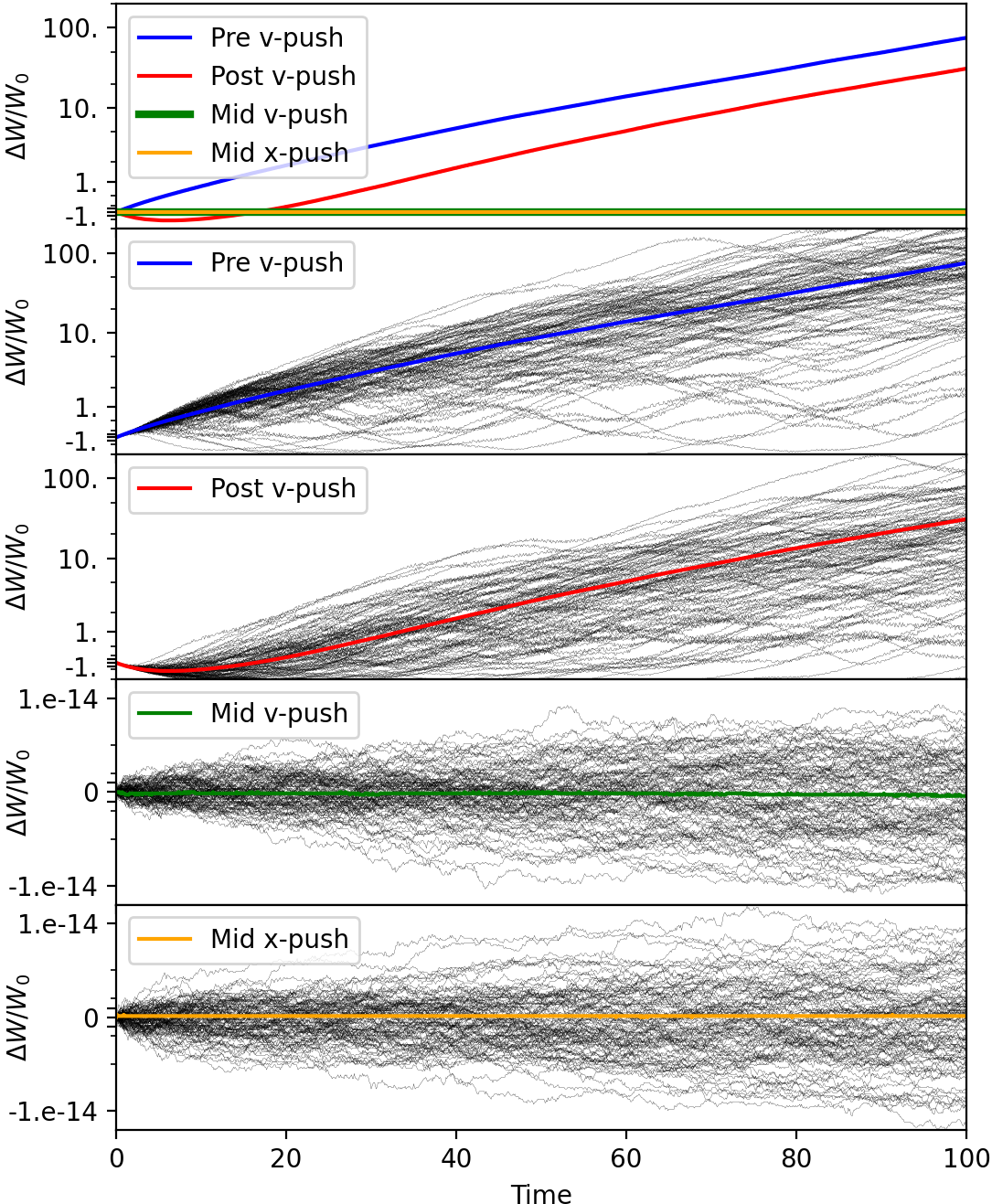}
\caption{(top) Average relative total energy deviation for an ensemble of 1000 non-interacting particles in a 3D harmonic oscillator experiencing transverse velocity kicks at every time steps, with MCC at pre v-push (blue), post v-push (red), mid v-push (green) or mid x-push (orange) location. The four bottom plots overlay the average of the 1000 relative energy deviation history to a sample of 100 histories for each case. 
Eq.~(\ref{Eq:WV1/2}):$W^{n+1/2}_{(1)}$ was used to compute the energy for the pre, post and mid v-push algorithms. 
Eq.~(\ref{Eq:WV}):$W^{n}_{(4)}$ was used to compute the energy for the mid x-push algorithm. 
}
\label{fig:HOaverage} 
\end{figure}

The numerical experiment was then extended to an ensemble of 1000 non-interacting particles evolving in a 3D harmonic oscillator while experiencing random transverse velocity kicks at every time steps. The evolution of the relative energy deviation as a function of time is plotted in Fig.~\ref{fig:HOaverage} for MCC at pre v-push (blue), post v-push (red), mid v-push (green) or mid x-push (orange) location. 
It shows that when placing the random transverse velocity kicks before the velocity push (pre v-push), the ensemble of particles is experiencing numerical heating, on average. 
Eq.~(\ref{Eq:WV1/2}):$W^{n+1/2}_{(1)}$ was used to compute the energy for the pre, post and mid v-push algorithms. 
Eq.~(\ref{Eq:WV}):$W^{n}_{(4)}$ was used to compute the energy for the mid x-push algorithm. 
When placing the random transverse velocity kicks after the velocity push (post v-push), the ensemble of particles is experiencing on average an initial period of numerical cooling that is followed by steady numerical heating at the same rate as with the pre v-push option. The new options of placing the random transverse velocity kicks in the middle of the velocity or the position push lead to exact energy conservation to machine precision.
While the plots are not shown for brevity, it was verified that the overall conclusions, that there is secular growth of the total energy with the pre v-push and post v-push options while there is not with the mid v-push and mid x-push options, holds for every of the four expressions of discrete energy from Eqs.~(\ref{Eq:WV1/2})-(\ref{Eq:WV}).

The exploration of the simple model with one particle with random transverse velocity kicks while experiencing harmonic oscillations shows that the numerical heating observed in full PIC simulations can be recovered without invoking collective, interpolation or radiative effects. 
It also confirms that the proposed algorithm, with proper centering of the MCC module 
in the middle of the velocity or position push, mitigates the issue, with no numerical heating, and with exact energy conservation for this simple model.
The confirmation that the main conclusions of the exploration of this simple model extend to the full PIC algorithm is given in the next section using simulations and analysis.

\subsection{\label{subsec:Analysis_fullPIC}Analysis of energy conservation for the full PIC-MCC cycle}

In section~\ref{subsec:HO}, we showed that the proposed mid v-push and mid x-push algorithm conserve energy \emph{exactly} for a simplified, single-particle model where:
\begin{itemize}
\item the electromagnetic PIC interaction between particles is replaced by a fixed, single-particle harmonic oscillator potential;
\item MCC collisions are replaced by random transverse velocity kicks. 
\end{itemize}

This section generalizes this result for the full PIC cycle with MCC binary collisions, confirming that the properties observed and analyzed with a single particle are preserved when including interpolations between macroparticles and gridded quantities, as well as collective and radiative effects. 
A reminder of the energy-conserving properties of the explicit PIC algorithm \emph{without} collisions (Sec.~\ref{subsubsec:analysis_no_collisions}) is given first. Unlike the harmonic oscillator model, the explicit PIC cycle is only exactly energy conserving in the asymptotic limit of vanishing time steps. It is then shown (Sec.~\ref{subsubsec:analysis_prepostvpush}) that incorporating MCC collisions with the post/pre v-push algorithm results in an additional term in the energy balance equation, which leads to spurious heating, as discussed also in \cite{AlvesPRE2021}.
By contrast, the mid v-push algorithm does not result in such a term, and thus leaves the energy conservation properties unchanged (Sec.~\ref{subsubsec:analysis_midvpush}). 
The case of the mid-x push algorithm happens to be more complex, and does not leave the energy balance equation unchanged. For this reason, the analysis of energy conservation for the mid x-push algorithm is discussed in Appendix~\ref{app:analysis_midxpush}.

\subsection{\label{subsubsec:analysis_no_collisions}Energy conservation without collisions}

The standard explicit leap-frog electromagnetic PIC method can be written as
\begin{align}
\text{Stage 1:} \ \ \ \ &  \frac{\Bb_g^{n+1/2}-\Bb_g^{n-1/2}}{\Delta t} = -\nabla\times\Eb_g^{n}, \label{BgEq_fullPIC} \\
\text{Stage 2:} \ \ \ \ &   m_p\frac{\vb^{n+1/2}_p-\vb_p^{n-1/2}}{\Delta t} = q_p\left(\Eb_p^n+\frac{\vb_p^{n+1/2}+\vb_p^{n-1/2}}{2}\times\Bb_p^{n}\right), \label{vpEq_fullPIC} \\
\text{Stage 3:} \ \ \ \ &   \frac{\xb_p^{n+1}-\xb_p^n}{\Delta t} = \vb_p^{n+1/2}, \label{xpEq_fullPIC} \\
\text{Stage 4:} \ \ \ \ &  \frac{\Eb_g^{n+1}-\Eb_g^n}{c^2\Delta t} = \nabla\times\Bb_g^{n+1/2} - \mu_0\sum_p\frac{q_p}{\Delta V}S^{n+1/2}_{gp}\vb_p^{n+1/2}, \label{EgEq_fullPIC}
\end{align}
where $\Delta V$ is the volume of a cell, and where $\Eb_p^n = \sum_gS^n_{gp}\Eb_g^n$ 
and $\Bb_p^n = \sum_gS^n_{gp}\left(\Bb_g^{n+1/2}+\Bb_g^{n-1/2}\right)/2$ are the fields gathered from the grid points $g$ to the particle $p$ using the shape factor $S_{gp}^n\equiv S_{g}(\textbf{x}_p^n)$. When using direct deposition \cite{Morsenielson1971}, the shape factor used to deposit the particle current to the grid in Eq.~(\ref{EgEq_fullPIC}) is at time $t_{n+1/2}$ and can be expressed as $S^{n+1/2}_{gp}\equiv S^{n+1/2}_{g}(\xb_p^{n+1/2})$.
Relativistic effects are ignored for simplicity.\\

The particle energy law is obtained by taking the dot product of Eq.~(\ref{vpEq_fullPIC}) with $\left(\vb^{n+1/2}_p+\vb^{n-1/2}_p\right)/2$ and summing over all particles:
\begin{align}
  \sum_p\frac{\mathcal{E}_p^{n+1/2}-\mathcal{E}_p^{n-1/2}}{\Delta t} = \sum_pq_p\frac{\vb_p^{n+1/2} + \vb_p^{n-1/2}}{2}\cdot\Eb_p^n, \label{Eq:Eparts_fullPIC}
\end{align}
where $\mathcal{E}_p^{n+1/2}\equiv m_p|\vb_p^{n+1/2}|^2/2$ is the kinetic energy of particle $p$ at time $t_{n+1/2}$. (Note that this expression is identical to the kinetic energy term in $W_{(1)}^{n+1/2}$, in Eq.~(\ref{Eq:WV1/2}).) The energy law for the fields is obtained by taking the dot product of Eq.~(\ref{BgEq_fullPIC}) with $\Delta V\left(\Bb_g^{n+1/2}+\Bb_g^{n-1/2}\right)/(2\mu_0)$, Eq.~(\ref{EgEq_fullPIC}) with $\epsilon_0\Delta V\Eb_g^n/2$, and Eq.~(\ref{EgEq_fullPIC}) shifted in time backward by one time step also with $\epsilon_0\Delta V\Eb_g^n/2$, summing these three expressions, and then summing over all grid points. This gives
\begin{align}
   \nonumber \sum_g\frac{\mathcal{E}_g^{n+1/2}-\mathcal{E}_g^{n-1/2}}{\Delta t} &+ \frac{1}{\mu_0}\sum_g\left(\Bb_g^n\cdot\nabla\times\Eb_g^{n} - \Eb_g^{n}\cdot\nabla\times\Bb_g^n \right)\Delta V \\
    &= -\sum_g\Eb_g^n\cdot\sum_pq_p\frac{S^{n+1/2}_{gp}\vb_p^{n+1/2}+S^{n-1/2}_{gp}\vb_p^{n-1/2}}{2}, \label{Eq:Efields_fullPIC}
\end{align}
where the energy on the grid is a sum of energy in the electric and magnetic fields: $\mathcal{E}^{n+1/2}_g\equiv \epsilon_0\Delta V\Eb^n_g\cdot\Eb_g^{n+1}/2 + \Delta V|\Bb^{n+1/2}|^2/(2\mu_0)$. The second term on the left-hand-side of Eq.~(\ref{Eq:Efields_fullPIC}) corresponds the Poynting flux term at the boundaries of the simulation domain and thus can be dropped for this analysis, which does not include energy variations from loss or injection of fields or particles at the boundaries.

Defining $W^{n+1/2}_{tot}\equiv \sum_g\mathcal{E}_g^{n+1/2}+\sum_p\mathcal{E}_p^{n+1/2}$ to be the total energy in the fields and particles at time $t_{n\pm 1/2}$. The total energy law for the full PIC system of equations given in Eqs.~(\ref{BgEq_fullPIC})-(\ref{EgEq_fullPIC}) is obtained by adding Eqs.\,(\ref{Eq:Eparts_fullPIC}) and (\ref{Eq:Efields_fullPIC}) together, ving
\begin{align}
   \frac{W_{tot}^{n+1/2}-W_{tot}^{n-1/2}}{\Delta t} =- \sum_g\sum_pq_p\Eb_g^n\cdot\frac{\left(S^{n+1/2}_{gp}-S^n_{gp}\right)\vb_p^{n+1/2}+\left(S^{n-1/2}_{gp}-S_{gp}^n\right)\vb_p^{n-1/2}}{2}. \label{Eq:Etot_fullPIC}
\end{align}
In Appendix \ref{app:taylor_expansion}, it is shown that the right-hand-side (RHS) of Eq.~(\ref{Eq:Etot_fullPIC}) tends to zero in the limit of vanishing time steps. The next section examines how the RHS of Eq.~(\ref{Eq:Etot_fullPIC}) is modified when incorporating MCC collisions in the PIC cycle.

\subsection{\label{subsubsec:analysis_prepostvpush}Pre v-push and post v-push collisions}

With the pre v-push method, Stage 2 in Eq.~(\ref{vpEq_fullPIC}) is replaced with the following two steps:
\begin{align}
\text{Stage 2a:} \ \ \ \ &   \vb_p^{n-1/2}\rightarrow \tilde{\vb}_p^{n-1/2}, \\
\text{Stage 2b:} \ \ \ \ &   m_p\frac{\vb^{n+1/2}_p-\tilde{\vb}_p^{n-1/2}}{\Delta t} = q_p\left(\Eb_p^n+\frac{\vb_p^{n+1/2}+\tilde{\vb}_p^{n-1/2}}{2}\times\Bb_p^{n}\right), \label{vpEq_PIC-MCC_prePush}
\end{align}
where $\tilde{\vb}_p$ represents particle velocities post-collision. The collision operator is assumed to be energy conserving such that $\sum_p\tilde{\mathcal{E}}_p = \sum_p\mathcal{E}_p$, and the energy law for the particles becomes
\begin{align}
  \text{pre v-push collision}: \ \ \ \sum_p\frac{\mathcal{E}_p^{n+1/2}-\mathcal{E}_p^{n-1/2}}{\Delta t} = \sum_pq_p\frac{\vb_p^{n+1/2} + \tilde{\vb}_p^{n-1/2}}{2}\cdot\Eb_p^n. \label{Eq:Eparts_PIC-MCC_prePush}
\end{align}
The only difference between the particle energy law here and that without collision given in Eq.~(\ref{Eq:Eparts_fullPIC}) is that $\vb_p^{n-1/2}$ is replaced by the post-collision velocity $\tilde{\vb}_p^{n-1/2}$ in the RHS. This is a critical difference because the energy law for the fields given in Eq.~(\ref{Eq:Efields_fullPIC}) is not altered by collisions for the pre v-push method. The total energy law for full PIC with pre v-push collision is
\begin{align}
\frac{W_{tot}^{n+1/2}-W_{tot}^{n-1/2}}{\Delta t} =& - \sum_{g} \sum_p q_p\Eb_g^n\cdot\frac{\left(S^{n+1/2}_{gp}-S^n_{gp}\right)\vb_p^{n+1/2}+\left(S^{n-1/2}_{gp}-S_{gp}^n\right)\vb_p^{n-1/2}}{2} \nonumber\\
&+ \sum_{g} \sum_p q_p\Eb_g^n\cdot\delta \vb_p^{n-1/2}\frac{S^n_{gp}}{2}. \label{Eq:Etot_PIC-MCC_prePush}
\end{align}
where $\delta \vb_p^{n-1/2} \equiv \tilde{\vb}_p^{n-1/2}-\vb_p^{n-1/2}$ is the change of velocity due to collisions.

The first term in the RHS is identical to that in Eq.~(\ref{Eq:Efields_fullPIC}), but the second term (involving $\delta \vb_p^{n-1/2}$) is a new term. In Ref.~\cite{AlvesPRE2021}, it is shown that the amplitude of this second term scales as $\nu (\omega_p\Delta t)^2$, where $\nu$ is the characteristic
frequency of Coulomb collisions, and that it is precisely this term that results in the net energy gain of the system over time. This is consistent with the growth of total energy seen in the bottom left panel of Fig.~\ref{fig:energy_standard_pic_mcc}, which uses the pre v-push algorithm.

If the collision is applied post v-push rather than pre v-push, then Stage 2 in Eq.~(\ref{vpEq_fullPIC}) is replaced with the following two steps:
\begin{align}
\text{Stage 2a:} \ \ \ \ &   m_p\frac{\vb^{n+1/2}_p-\tilde{\vb}_p^{n-1/2}}{\Delta t} = q_p\left(\Eb_p^n+\frac{\vb_p^{n+1/2}+\tilde{\vb}_p^{n-1/2}}{2}\times\Bb_p^{n}\right), \label{vpEq_PIC-MCC_postPush} \\
\text{Stage 2b:} \ \ \ \ &   \vb_p^{n+1/2}\rightarrow \tilde{\vb}_p^{n+1/2}.
\end{align}
This results in the following total energy law:
\begin{align}
\frac{W_{tot}^{n+1/2}-W_{tot}^{n-1/2}}{\Delta t} = & - \sum_{g} \sum_p q_p\Eb_g^n\cdot\frac{\left(S^{n+1/2}_{gp}-S^n_{gp}\right)\vb_p^{n+1/2}+\left(S^{n-1/2}_{gp}-S_{gp}^n\right)\vb_p^{n-1/2}}{2} \nonumber\\
&-\sum_g\sum_pq_p\Eb_g^n\cdot\delta \vb_p^{n+1/2}\frac{S^n_{gp}}{2}. \label{Eq:Etot_PIC-MCC_postPush}
\end{align}
Again, there is an additional term compared to Eq.~(\ref{Eq:Efields_fullPIC}).
Separate numerical simulations (not shown here) revealed that the long time behavior of this method produces heating at a similar rate to the pre v-push method, in line with the analogous harmonic oscillator models discussed previously.

\subsection{\label{subsubsec:analysis_midvpush}Mid v-push collisions}
In the case of the mid v-push, the particle velocity advance (stage 2) is done in five substages as follows, where the Boris pusher is used to advance particles:
\begin{align}
 \text{Stage 2a:} \ \ \ \ &   m_p(\vb'_p-\vb_p^{n-1/2}) = q_p\Eb_p^n\frac{\Delta t}{2}, \label{vpEq_full_strang1} \\
 \text{Stage 2b:} \ \ \ \ &   m_p(\vb_p^*-\vb'_p) = q_p\frac{(\vb^*_p+\vb'_p)}{2}\times\Bb_p^n\frac{\Delta t}{2}, \label{vpEq_full_strang3} \\
\text{Stage 2c:} \ \ \ \ & \ \ \ \ \vb_p^{*} \rightarrow \tilde{\vb}_p^{*}, \label{scatter_full_strang3} \\
\text{Stage 2d:} \ \ \ \ &   m_p(\vb''_p-\tilde{\vb}_p^{*}) = q_p\frac{(\vb''_p+\tilde{\vb}^*_p)}{2}\times\Bb_p^n\frac{\Delta t}{2}, \label{vpEq_full_strang4} \\
  \text{Stage 2e:} \ \ \ \ &  m_p(\vb_p^{n+1/2}-\vb''_p)
    = q_p\Eb_p^n\frac{\Delta t}{2}, \label{vpEq_tvs_strang5}
\end{align}
This is the standard Boris push with the magnetic field rotation split into two half $\Delta t$ rotations with a full $\Delta t$ collision operation done in between. 

By dotting Eq.~(\ref{vpEq_tvs_strang5}) and Eq.~(\ref{vpEq_full_strang1}) with $(\vb^{n+1/2}_p + \vb''_p)/2$ and $(\vb'_p+\vb_p^{n-1/2})/2$ respectively, and by using the fact that magnetic rotation in stage 2b and 2d conserve the kinetic energy of each particle (i.e., $\mathcal{E}_p^*=\mathcal{E}_p'$, $\mathcal{E}_p''=\tilde{\mathcal{E}}_p^*$), and that the collisions in stage 2c conserves the total kinetic energy (i.e., $\sum_p \tilde{\mathcal{E}}_p^* = \sum_p \mathcal{E}_p^* $), the energy law for the particles after all five of these substages is
\begin{align}
   \nonumber \sum_p\frac{\mathcal{E}_p^{n+1/2}-\mathcal{E}_p^{n-1/2}}{\Delta t} &= \sum_pq_p\frac{(\vb_p^{n+1/2} + \vb''_p) + (\vb'_p+\vb_p^{n-1/2})}{4}\cdot\Eb_p^n, \\
    \nonumber &= \sum_pq_p\frac{(\vb_p^{n+1/2} + \vb_p^{n+1/2}+\frac{q_p\Eb^n_p\Delta t}{2m_p}) + (\vb^{n-1/2}_p -\frac{q_p\Eb^n_p\Delta t}{2m_p} + \vb_p^{n-1/2})}{4}\cdot\Eb_p^n, \\
    &= \sum_pq_p\frac{\vb_p^{n+1/2} + \vb_p^{n-1/2}}{2}\cdot\Eb_p^n, \label{Eq:Eptot_stage1to5}
\end{align}
where we used Eq.~(\ref{vpEq_tvs_strang5}) and Eq.~(\ref{vpEq_full_strang1}) again in the second line.

The above energy law for particles Eq.~(\ref{Eq:Eptot_stage1to5}) is identical to that in the absence of collisions given previously in Eq.~(\ref{Eq:Eparts_fullPIC}). Furthermore, since the rest of the PIC loop is unchanged for the mid v-push method, the field energy law Eq.~(\ref{Eq:Efields_fullPIC}) is also unchanged, and thus the total energy law for this method is identical to that in the absence of collision given in Eq.~(\ref{Eq:Etot_fullPIC}).

Thus, the mid v-push algorithm retains the energy conservation property of the collision-free algorithm mentioned in Sec.~\ref{subsubsec:analysis_no_collisions}, and in particular there is no additional growth of energy due to collisions. This is again consistent with the upper right panel of Fig.~\ref{fig:three graphs}.

\section{\label{sec:implementation}Discussion of practical implementation in a PIC code}

This section explores practical considerations that guide the actual implementation of the mid v-push and mid x-push options.

\subsubsection{\label{subsubsec:implementation_midvpush}Mid v-push PIC-MCC}
 
Placing the MCC module in the middle of the velocity push implies to:
\begin{enumerate}
    \item split the velocity push in two halves,
    \item either gather the electromagnetic field components from the grid to the macroparticles twice or store the values in auxiliary arrays. This is because the MCC module must loop on all the particle species that participate to the collisions, hence the fields that are gathered to push the velocity over the first half step before MCC cannot be stored in local memory for reuse during the second half velocity push.
\end{enumerate}
Ideally, the combination of the two half velocity pushes should lead to the same result as one push with the unsplit algorithm to machine precision if no collision occurs. This is trivial for the Boris pusher, where the magnetic rotation can be split in an arbitrary number of substeps without changing the algorithm by scaling the time step used for the magnetic rotation accordingly, or it is known for other pushers, e.g., by using Eqs.(11-13) from~\cite{Vaypop2008} for the two half velocity push for the ``Vay'' pusher~\cite{Vaypop2008}. However, it may not be straightforward for other pushers such as, e.g., the Higuera-Cary pusher~\cite{Higuera2017} for which a separation in two half steps is not straightforward.
Gathering the fields twice or creating auxiliary arrays to store them can incur additional computational costs that may not be negligible and must be taken into consideration.
\subsubsection{\label{subsubsec:implementation_midvpush}Mid x-push PIC-MCC}
\begin{figure}
    \centering
    \includegraphics[scale=0.55]{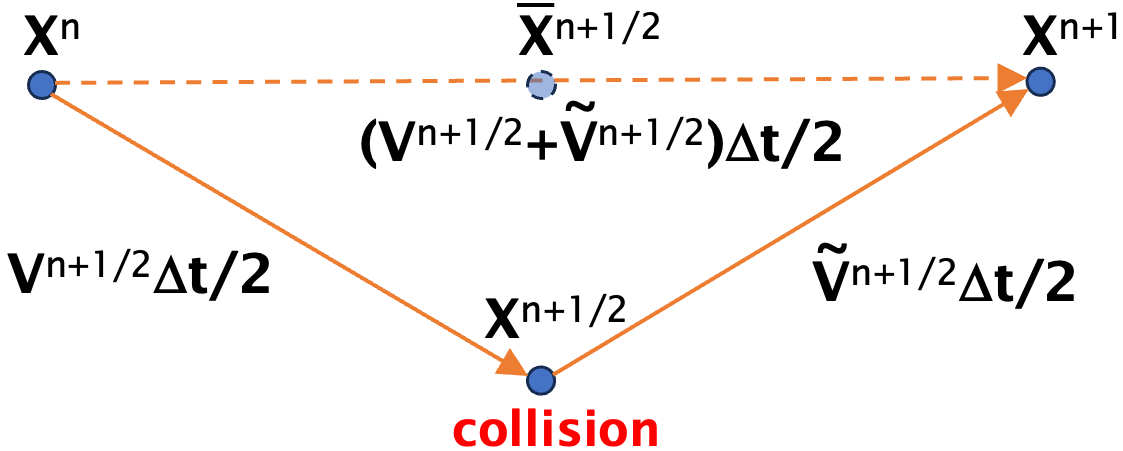}
    \caption{Diagram of the ``x-push'' MCC option, where the collision occurs in the middle of the position push, changing the velocity from $\vb^{n+1/2}$ to $\tilde{\vb}^{n+1/2}$. 
    The current deposition can be performed using either (a) the two half position push from $\xb^n$ to $\xb^{n+1/2}$ and $\xb^{n+1/2}$ to $\xb^{n}$, using $\vb^{n+1/2}$ and $\tilde{\vb}^{n+1/2}$, respectively, or (b) the average velocity $(\vb^{n+1/2}+\tilde{\vb}^{n+1/2})/2$ over the full time step.
    }
    \label{fig:xpush_diagram}
\end{figure}

Placing the MCC module in the middle of the position push implies to:
\begin{enumerate}
    \item split the position push in two halves,
    \item either deposit the current onto the grid twice using the velocities before and after collision, or depositing it once using the average of the two velocities, (see Fig.~\ref{fig:xpush_diagram}).
\end{enumerate}
The splitting of the position push in two halves is trivial since the position push is a simple linear operation. 
As illustrated in Fig.~\ref{fig:xpush_diagram}, the current deposition can be performed using either (a) the two half position push from $\xb^n$ to $\xb^{n+1/2}$ and $\xb^{n+1/2}$ to $\xb^{n}$, using $\vb^{n+1/2}$ and $\tilde{\vb}^{n+1/2}$, respectively, or (b) the average velocity $(\vb^{n+1/2}+\tilde{\vb}^{n+1/2})/2$ over the full time step. 
In either case, auxiliary arrays are needed to store either the velocity before the collision or the averaged velocity.

On von Neumann compute architectures used today, moving data between memory and the processing units accounts for a significant portion of the total computational time.
Algorithms that do not perform a minimum number of operations per Byte transferred are bound in their performance by memory bandwidth.
On contemporary CPU and GPU hardware implementations, the central routines of the particle-in-cell cycle (gather, push, scatter, field update), are all memory bandwidth bound. 
Hence depositing the current for the two half velocity push should not lead to a significant performance hit in comparison to depositing only once with the average velocity, provided that the two depositions happen in the same inner loop.

Both methods have performed equally with regard to energy conservation  on the 2D uniform plasma and the 1D magnetic piston tests reported here. 
The trade-offs between the two-velocity deposition method being better for accuracy (as could be assumed because it follows the physics more closely) and the one-average-velocity deposition being better for efficiency (as can be assumed because it involves less computational operations) will need more testing and analysis, taking also into account the extra memory cost for storing extra components for particles (electromagnetic fields or velocities), which are left for future work. 


\section{\label{sec:conclusion}Conclusion}

This paper elucidated the origin of an anomalous numerical heating that was observed when coupling the PIC and MCC techniques in the explicit PIC-MCC algorithm for both electrostatic and electromagnetic PIC-MCC.
It was observed that the standard implementations of PIC-MCC were generally not time-centered and two placements of MCC in the PIC loop were proposed and studied: in the middle of the velocity push or in the middle of the position push.
It was shown on 2D periodic plasma simulations that centering the MCC events properly in the PIC loop, as proposed, prevents the anomalous numerical heating that occurs otherwise.
Furthermore, an example of application to the modeling of a magnetic-driven piston collisional shock simulations showed a dramatic reduction of numerical heating when using the new proposed placement of MCC in the PIC loop.
Next, a simple analysis of a single particle experiencing harmonic oscillations showed that the numerical heating and its prevention can be obtained and analyzed without invoking any collective, interpolation or radiative effects that can occur in electromagnetic PIC-MCC simulations.
It was then shown that these properties do persist in the analysis of the full electromagnetic PIC-MCC loop. 
Future work includes the study of the practical advantages and disadvantages of the various implementation options that were proposed in the paper, as well as the exploration of the benefits of the proposed scheme to a wider range of applications.

\begin{acknowledgments}

The authors acknowledge seminal discussions with Brendan Godfrey, who expressed doubts about previous work, thus motivating the work presented in this paper.

This research used the open-source particle-in-cell code WarpX https://github.com/ECP-WarpX/WarpX. Primary WarpX contributors are with LBNL, LLNL, CEA-LIDYL, SLAC, DESY, CERN, and TAE Technologies. We acknowledge all WarpX contributors.
This material is based upon work supported by the KISMET collaboration, a project of the U.S. Department of Energy, Office of Science, Office of Advanced Scientific Computing Research and Office of High Energy Physics, Scientific Discovery through Advanced Computing (SciDAC) program.
This research used resources of the National Energy Research Scientific Computing Center, a DOE Office of Science User Facility supported by the Office of Science of the U.S. Department of Energy under Contract No. DE-AC02-05CH11231 using NERSC award FES-ERCAP0027617.
Work supported by a US Department of Energy under contract DE-AC52-07NA27344.LLNL-ABS-863974.

J.-L. V. conceptualized the main concept, performed initial proof-of-concept validation with the PIC code Warp, realized the initial implementation of the method in the PIC code WarpX, developed the analysis of the single-particle harmonic oscillator model, wrote the corresponding section, and led the research and writing of the paper. 
J. A. performed simulations with the new method using the PIC code PICNIC for exploration and verification, led the formal analysis of the full PIC-MCC cycle and the writing of the corresponding section. 
O. S. assisted the implementation of the method in WarpX, conducted the WarpX simulations for the 2-D uniform plasma and the 1-D piston cases, and contributed to the corresponding sections.
R. L. assisted in the formal analysis of the full PIC-MCC cycle and participated to the organization, review and writing of the paper.
D. G. assisted in the preparation of the simulations of the magnetic piston and contributed to the discussions, the review and writing of the paper.
A. H. contributed to the performance analysis, discussions and to the review and writing of the paper.
\end{acknowledgments}

\appendix
\counterwithin{equation}{section}

\section{Energy conservation with the harmonic oscillator model}
\label{sec:Appendix_HO_energy_conservation}
\numberwithin{equation}{section}

This appendix describes the derivations of the expressions of discrete energy that conserve energy to machine precision for one particle experiencing harmonic oscillations.

\subsection{Energy conservation without transverse velocity kicks}
\label{subsec:Appendix_HO_energy_conservation_nocollision}

Eq.~(\ref{Eq:WV1/2}) is obtained by multiplying (\ref{Eq:HOV}) by $(\vb^{n+1/2}+\vb^{n-1/2})$, plugging (\ref{Eq:HOE}) and replacing $\vb^{n+1/2}$ and $\vb^{n-1/2}$ using (\ref{Eq:HOX}), then rearranging to identify the same expression for the energy at two consecutive time steps. 
Eq.~(\ref{Eq:WX}) is obtained very similarly by multiplying (\ref{Eq:HOX}) by $(\xb^{n+1}+\xb^n)$, then following the same steps, replacing $\xb^{n+1}$ and $\xb^n$ on the right hand side using (\ref{Eq:HOV}) and (\ref{Eq:HOE}) and rearranging. 
Eqs.~(\ref{Eq:WX1/2}) and (\ref{Eq:WV}) are obtained in a very similar manner using
\begin{align}
    \vb^{n+1} - \vb^{n} &= -\frac{q\kappa}{m} \xb^{n+1/2} \Delta t,\\
    \xb^{n+1/2} - \xb^{n-1/2} &= \vb^{n} \Delta t, \label{Eq:HOXmid}
\end{align}
which follow from Eqs.~(\ref{Eq:HOV}-\ref{Eq:HOE}).

\subsection{Energy conservation with mid v-push MCC}
\label{subsec:Appendix_HO_energy_conservation_midvpush}

Denoting $\delta\vb$ the transverse velocity kicks, the update on the velocity becomes:
\begin{align}
    \vb^{n+1/2} - \vb^{n-1/2} &= -\frac{q\kappa}{m} \xb^n \Delta t + \delta\vb, \label{Eq:HOVwcol}
\end{align}
while the equation on the position update is unchanged. 
This can be rewritten
\begin{align}
    \vb^{n} &= \vb^{n-1/2} -\frac{q\kappa}{2m} \xb^n \Delta t, \label{Eq:HO_vpush_vn}\\
    \tilde{\vb}^n &= \vb^n + \delta\vb, \\
    \vb^{n+1/2} &= \tilde{\vb}^n -\frac{q\kappa}{2m} \xb^n \Delta t \label{Eq:HO_vpush_vnp5}
\end{align}

The conservation of energy before and after the transverse velocity kick implies that
$(\vb^n)^2 = (\vb^n + \delta\vb)^2 =(\tilde{\vb}^n)^2 = (\tilde{\vb}^n - \delta\vb)^2$.
 This implies, when combined with Eqs.~(\ref{Eq:HO_vpush_vn})-(\ref{Eq:HO_vpush_vnp5}), that $\delta\vb(\vb^{n-1/2}+\vb^{n+1/2})=0$, which, when multiplying Eq.~(\ref{Eq:HOVwcol}) by $(\vb^{n+1/2}+\vb^{n-1/2})$ leads to the same relation for conservation of energy as without the transverse velocity kick, hence to Eq.~(\ref{Eq:WV1/2}):$W^{n+1/2}_{(1)}$.

\subsection{Energy conservation with mid x-push MCC}
\label{subsec:Appendix_HO_energy_conservation_midxpush}
In this case, the leapfrog loop becomes
\begin{align}
    \vb^{n+1/2} &= \vb^{n} - \frac{\kappa q}{2m} \xb^n \Delta t,  \label{Eq:Verlet1_midx}\\
    \xb^{n+1/2} &= \xb^n + \vb^{n+1/2} \Delta t/2, \label{Eq:Verlet2_midx}\\
    \vb^{n+1/2} &\rightarrow \tilde{\vb}^{n+1/2} \label{Eq:Verlet_midx_collision}\\
    \xb^{n+1} &= \xb^{n+1/2} + \tilde{\vb}^{n+1/2} \Delta t/2,  \label{Eq:Verlet3_midx}\\
    \vb^{n+1} &= \tilde{\vb}^{n+1/2}- \frac{\kappa q}{2m} \xb^{n+1} \Delta t.  \label{Eq:Verlet4_midx}
\end{align}
where Eq.~(\ref{Eq:Verlet_midx_collision}) represents the transverse velocity kick.

The change in kinetic energy between times $n$ and $n+1$ is given by:
\begin{align}
\frac{m}{2}(\vb^{n+1})^2 - \frac{m}{2}(\vb^n)^2 &= \frac{m}{2}\left( \tilde{\vb}^{n+1/2}-\frac{\kappa q \Delta t}{2m} \xb^{n+1}\right)^2 - \frac{m}{2}\left( \vb^{n+1/2} + \frac{\kappa q \Delta t}{2m} \xb^{n}\right)^2 \\
&= \frac{m}{2}\left( \left(1-\frac{\kappa q \Delta t^2}{4m}\right)\tilde{\vb}^{n+1/2}-\frac{\kappa q \Delta t}{2m} \xb^{n+1/2}\right)^2 \nonumber \\
&\qquad - \frac{m}{2}\left( \left(1-\frac{\kappa q \Delta t^2}{4m}\right)\vb^{n+1/2} + \frac{\kappa q \Delta t}{2m} \xb^{n+1/2}\right)^2 \\
&= \left(1-\frac{\kappa q \Delta t^2}{4m}\right)^2\left[ \frac{m}{2}(\tilde{\vb}^{n+1/2})^2 - \frac{m}{2} (\vb^{n+1/2})^2\right] \nonumber \\
&\qquad - \frac{\kappa q \Delta t}{2}\left(1-\frac{\kappa q \Delta t^2}{4m}\right)\xb^{n+1/2}\cdot(\vb^{n+1/2}+\tilde{\vb}^{n+1/2})
\label{Eq:MidxLastLineExpanded}
\end{align}
where Eq.~(\ref{Eq:Verlet4_midx}) and Eq.~(\ref{Eq:Verlet1_midx}) were used to replace $\vb^{n+1}$ and $\vb^n$ in the first line, while Eq.~(\ref{Eq:Verlet3_midx}) and Eq.~(\ref{Eq:Verlet2_midx}) were used to replace $\xb^{n+1}$ and $\xb^n$ in the second line. In the third line, the squared parentheses were expanded and similar terms were regrouped. 

The conservation of energy of the transverse velocity kick implies that $m(\tilde{\vb}^{n+1/2})^2/2 = m(\vb^{n+1/2})^2/2$, and thus the square bracket in Eq.~(\ref{Eq:MidxLastLineExpanded}) cancels out. 
In order to rewrite the second term in Eq.~(\ref{Eq:MidxLastLineExpanded}), we note that:
\begin{align}
\xb^{n+3/2}-\xb^{n-1/2} &= \vb^{n+3/2}\Delta t/2 + \xb^{n+1} - \xb^n + \tilde{\vb}^{n-1/2}\Delta t/2\\
&= \left( \tilde{\vb}^{n+1/2} -\frac{\kappa q}{m}\xb^{n+1}\Delta t \right)\frac{\Delta t}{2} + \xb^{n+1} - \xb^{n} + \left( \vb^{n+1/2} + \frac{\kappa q}{m}\xb^n \Delta t\right) \frac{\Delta t}{2}\\
&= (\tilde{\vb}^{n+1/2}+\vb^{n+1/2})\frac{\Delta t}{2}
+ \left( 1 - \frac{\kappa q \Delta t^2}{2m}\right)(\xb^{n+1}-\xb^n) \\
&= (\tilde{\vb}^{n+1/2}+\vb^{n+1/2})\frac{\Delta t}{2}
+ \left( 1 - \frac{\kappa q \Delta t^2}{2m}\right)(\tilde{\vb}^{n+1/2}+\vb^{n+1/2})\frac{\Delta t}{2} \\
&= \left( 1 - \frac{\kappa q \Delta t^2}{4m}\right)(\tilde{\vb}^{n+1/2}+\vb^{n+1/2})\Delta t
\end{align}
where Eq.~(\ref{Eq:Verlet3_midx}) and Eq.~(\ref{Eq:Verlet2_midx}) were used to replace $\xb^{n+3/2}$ and $\xb^{n-1/2}$ in the first line, and Eq.~(\ref{Eq:Verlet4_midx}) and Eq.~(\ref{Eq:Verlet1_midx}) were used to replace $\vb^{n+3/2}$ and $\tilde{\vb}^{n-1/2}$ in the second line. Similar terms were regrouped in the third line, while Eq.~(\ref{Eq:Verlet3_midx}) and Eq.~(\ref{Eq:Verlet2_midx}) were used again in the fourth line to replace $\xb^{n+1}$ and $\xb^{n}$.

Inserting the above equality in Eq.~(\ref{Eq:MidxLastLineExpanded}) gives:
\begin{align}
\frac{m}{2}(\vb^{n+1})^2 - \frac{m}{2}(\vb^n)^2 &= -\frac{\kappa q}{2}\xb^{n+1/2}\cdot(\xb^{n+3/2}-\xb^{n-1/2})\\
\frac{m}{2}(\vb^{n+1})^2 + \frac{\kappa q}{2}\xb^{n+1/2}\cdot\xb^{n+3/2} &= \frac{m}{2}(\vb^n)^2+\frac{\kappa q}{2}\xb^{n+1/2}\cdot\xb^{n-1/2}
\end{align}


i.e., the same expression as Eq.~(\ref{Eq:WV}):$W^{n}_{(4)}$ for the conservation of energy.

\section{\label{app:taylor_expansion}Conservation of energy for the full PIC algorithm in the limit of vanishing time step}

As discussed in Sec.~\ref{subsec:Analysis_fullPIC}, the energy balance for the full PIC algorithm with periodic boundaries and in the non-relativistic approximation reads
\begin{align}
   \frac{W_{tot}^{n+1/2}-W_{tot}^{n-1/2}}{\Delta t} =- \sum_g\sum_pq_p\Eb_g^n\cdot\frac{\left(S^{n+1/2}_{gp}-S^n_{gp}\right)\vb_p^{n+1/2}+\left(S^{n-1/2}_{gp}-S_{gp}^n\right)\vb_p^{n-1/2}}{2}.
\end{align}

In the limit of vanishing time step, this can be simplified using a first-order Taylor expansion in $\Delta t$ for the shape factor terms:
\begin{align}
    S^{n\pm 1/2}_{gp}  - S_{gp}^n 
    &= S_g(\xb_p^{n\pm1/2}) - S_g(\xb_p^n)\\
    &= S_g(\xb_p^{n}\pm\vb_p^{n\pm1/2}\Delta t/2) - S_g(\xb_p^n)\\
    &= \pm (\vb_p^{n\pm1/2}\cdot\boldsymbol{\nabla})S_g(\xb_p^{n})\Delta t/2 \label{Eq:Taylor1D}
\end{align}
Note that a key element for this to hold is that we assumed that the shape factor for field gathering ($S^n_{gp}$) and current deposition ($S^{n+1/2}_{gp}$) have the same expression as a function of the particle position ($S_g(\xb)$).


Replacing these expressions in the energy balance equation, we have:
\begin{equation}
   \frac{W_{tot}^{n+1/2}-W_{tot}^{n-1/2}}{\Delta t} =- \frac{\Delta t}{2}\sum_g\sum_pq_p\Eb_g^n\cdot\frac{(\vb_p^{n+1/2}\cdot\boldsymbol{\nabla})S_g(\xb_p^{n})\vb_p^{n+1/2}-(\vb_p^{n-1/2}\cdot\boldsymbol{\nabla})S_g(\xb_p^{n})\vb_p^{n-1/2}}{2}. \label{Wtot-taylor-expanded}
\end{equation}

Moreover, from the equation for the velocity update Eq.~(\ref{vpEq_fullPIC}), one has $\vb_p^{n+1/2} = \vb_p^{n-1/2} + O(\Delta t)$. Replacing $\vb^{n+1/2}$ accordingly in Eq.~(\ref{Wtot-taylor-expanded}), it can be seen that the first-order term in $\Delta t$ in the RHS cancels out, and thus:
\begin{equation}
\frac{W_{tot}^{n+1/2}-W_{tot}^{n-1/2}}{\Delta t} = O(\Delta t^2),
\end{equation}
i.e., the error in energy conservation tends to zero (quadratically) in the limit of vanishing timestep.

\section{\label{app:analysis_midxpush}full PIC-MCC with mid x-push method for collision}

Finally, we consider the mid x-push method where the collision is done in the middle of the x-push
stages. For simplicity, we neglect the change of position in the particles shape over one timestep, i.e. $S_{gp}^n \approx S_{gp}^{n+1/2} \approx S_{gp}^{n-1/2}$. Lifting this assumption results in additional terms of the form Eq.~(\ref{Eq:Taylor1D}) that already appear in the absence of collisions, and are therefore not specific to the mid x-push algorithm. In this case, the PIC-MCC loop can be expressed as:
\begin{align}
\text{Stage 2:} \ \ \ \ &   m_p\frac{\vb^{n+1/2}_p-\tilde{\vb}_p^{n-1/2}}{\Delta t} = q_p\left(\Eb_p^n+\frac{\vb_p^{n+1/2}+\tilde{\vb}_p^{n-1/2}}{2}\times\Bb_p^{n}\right), \label{vpEq_fullPIC_xpush} \\
\text{Stage 3a:} \ \ \ \ &  \frac{\xb^{n+1/2}-\xb^n}{\Delta t} = \frac{\vb^{n+1/2}}{2},\\
\text{Stage 3b:} \ \ \ \ &   \vb_p^{n+1/2}\rightarrow \tilde{\vb}_p^{n+1/2},\\
\text{Stage 3c:} \ \ \ \ &  \frac{\xb^{n+1}-\xb^{n+1/2}}{\Delta t} = \frac{\tilde{\vb}^{n+1/2}}{2},\\
\text{Stage 4:} \ \ \ \ &  \frac{\Eb_g^{n+1}-\Eb_g^n}{c^2\Delta t} = \nabla\times\Bb_g^{n+1/2} - \mu_0\sum_p\frac{q_p}{\Delta V}S^{n}_{gp}\frac{\vb_p^{n+1/2}+\tilde{\vb}_p^{n+1/2}}{2}. \label{EgEq_fullPIC_xpush}
\end{align} 
The particle energy law for this method is the same as for the post v-push and pre v-push methods given in Eq.~(\ref{Eq:Eparts_PIC-MCC_prePush}). The field energy law, however, now takes the following form:
\begin{align}
   \nonumber \sum_g\frac{\mathcal{E}_g^{n+1/2}-\mathcal{E}_g^{n-1/2}}{\Delta t} &= -\sum_g\sum_pq_pS^n_{gp}\Eb_g^n\cdot\left(\frac{\vb_p^{n-1/2}+\tilde{\vb}_p^{n-1/2}+\vb_p^{n+1/2}+\tilde{\vb}_p^{n+1/2}}{4}\right), \\
   \nonumber &= -\sum_g\sum_pq_pS^n_{gp}\Eb_g^n\cdot\left(\frac{\tilde{\vb}_p^{n-1/2}+\vb_p^{n+1/2}}{2}\right) \\
   &- \sum_g\sum_pq_pS^n_{gp}\left[\Eb_g^n\cdot\frac{\left(\tilde{\vb}_p^{n+1/2}-\vb_p^{n+1/2}\right) - \left(\tilde{\vb}_p^{n-1/2}-\vb_p^{n-1/2}\right)}{4}\right]. \label{Eq:Efields_PIC-MCC_xpush}
\end{align}
The first term on the RHS in the final expression in Eq.~(\ref{Eq:Efields_PIC-MCC_xpush}) is equal and opposite to the source term for the particle energy law. However, the last term on the RHS is not identically zero and thus, in contrast to the v-push method, energy is not identically conserved in the asymptotic limit where the difference of particle shape factors over a time step tends to zero. This is also in contrast to the analogous mid x-push harmonic oscillator method, where energy is identically preserved for finite time steps. The difference between these two analogous models can be understood by first rewriting the term in square brackets in the last term on the RHS of Eq.~(\ref{Eq:Efields_PIC-MCC_xpush}) as follows:
\begin{align}
   \nonumber \Eb_g^n\cdot\frac{\left(\tilde{\vb}_p^{n+1/2}-\vb_p^{n+1/2}\right) - \left(\tilde{\vb}_p^{n-1/2}-\vb_p^{n-1/2}\right)}{4} &=  \\
   \nonumber \Eb_g^n\cdot\frac{\left(\tilde{\vb}_p^{n+1/2}-\vb_p^{n+1/2}\right)}{4} &- \Eb_g^{n-1}\cdot\frac{\left(\tilde{\vb}_p^{n-1/2}-\vb_p^{n-1/2}\right)}{4} \\ 
   -c^2\Delta t\left(\nabla\times\Bb_g^{n-1/2} - \mu_0\sum_{p'}\frac{q_{p'}}{\Delta V}S^{n-1}_{gp'}\frac{\tilde{\vb}^{n-1/2}_{p'}+\vb^{n-1/2}_{p'}}{2}\right)& \cdot\frac{\left(\tilde{\vb}_p^{n-1/2}-\vb_p^{n-1/2}\right)}{4}.
\end{align}

The first two terms on the RHS of this expression are of the form $\Eb_g^n\cdot\textbf{f}\left(t_{n+1/2}\right) - \Eb_g^{n-1}\cdot\textbf{f}\left(t_{n-1/2}\right)$, and thus it is possible to absorb these terms into a re-definition of the electric field energy as $\hat{\mathcal{E}}^{n+1/2}_{Eg} \equiv \epsilon_0\Delta V\hat{\Eb}_g^{n+1}\cdot\Eb_g^n/2$ where $\hat{\Eb}_g^{n+1}$ is a modified electric field defined as
\begin{equation}
    \hat{\Eb}_g^{n+1} = \Eb_g^{n+1} + \frac{\Delta t}{\epsilon_0}\sum_p\frac{q_p}{\Delta V}S_{gp}\frac{\tilde{\vb}^{n+1/2}_p-\vb^{n+1/2}_p}{2}.
\end{equation}
Adopting this modified field energy definition, the total energy law for the x-push method can be written as
\begin{align}
\nonumber &\sum_p\frac{\left(\mathcal{E}_p^{n+1/2}-\mathcal{E}_p^{n-1/2}\right)}{\Delta t} + \sum_g\frac{\left(\hat{\mathcal{E}}_g^{n+1/2}-\hat{\mathcal{E}}_g^{n-1/2}\right)}{\Delta t} \\
&= \sum_g\sum_pq_pS_{gp}^n\left[c^2\Delta t\left(\nabla\times\Bb_g^{n-1/2} - \mu_0\sum_{p'}\frac{q_{p'}}{\Delta V}S^{n-1}_{gp'}\frac{\tilde{\vb}^{n-1/2}_{p'}+\vb^{n-1/2}_{p'}}{2}\right)\cdot\frac{\left(\tilde{\vb}_p^{n-1/2}-\vb_p^{n-1/2}\right)}{4}\right]. \label{Eq:Etotal_PIC-MCC_xpush}
\end{align}
The expression on the RHS of this equation is still not identically zero, and thus energy is not identically conserved for the x-push method, but the error term has a different nature than that for the pre v-push and post v-push methods and the connection with the analogous harmonic oscillator model, where energy is identically conserved, can now be made. First, there is no magnetic field in the Harmonic oscillator and so the $\nabla\times\Bb_g^{n-1/2}$ term can be ignored. The main difference now between full PIC-MCC and the harmonic oscillator model is that the latter a) only considers a single particle and b) the collision operator is just a simple rotation of the velocity vector. For a velocity rotation of a single particle, the average of the post-scatter and pre-scatter velocities is orthogonal to the difference. That is, $\left(\tilde{\textbf{v}}^{n-1/2}_p+\textbf{v}^{n-1/2}_p\right)\cdot\left(\tilde{\textbf{v}}^{n-1/2}_p-\textbf{v}^{n-1/2}_p\right)=0$. For a single particle, and for a collision operator that is just a rotation of the velocity vector, the last term on the RHS of Eq.~(\ref{Eq:Etotal_PIC-MCC_xpush}) is identically zero, consistent with the harmonic oscillator model.

While the RHS of Eq.~(\ref{Eq:Etotal_PIC-MCC_xpush}) is identically zero for a single particle in an unmagnetized system with a simplified collision operator, it is not identically zero for an arbitrary number of particles undergoing energy- and momentum-preserving collision of binary pairs of particles. This is in contrast with the v-push model for full PIC-MCC, where exact energy conservation is obtained analytically in the asymptotic limit where the difference in particle shape factors over a time step tends to zero. The proof of energy-conservation for the x-push model lies in the fact that the RHS of Eq.~(\ref{Eq:Etotal_PIC-MCC_xpush}) does not grow in time. A formal proof of this is beyond the scope of this paper. Instead, numerical simulation results using this algorithm for full PIC-MCC simulations of various cases are used to illustrate that numerical heating does not occur.

\bibliography{Mendeley_Biblio_Vay}

\end{document}

%% file: newcommands.tex
\newcommand{\ub}{\boldsymbol{u}}
\newcommand{\vb}{\boldsymbol{v}}
\newcommand{\xb}{\boldsymbol{x}}

\newcommand{\Bb}{\boldsymbol{B}}

\newcommand{\Eb}{\boldsymbol{E}}

\DeclareMathAlphabet\mathbfcal{OMS}{cmsy}{b}{n}

